\documentclass[fleqn,usenatbib]{mnras}

\usepackage{newtxtext,newtxmath}

\usepackage[T1]{fontenc}

\DeclareRobustCommand{\VAN}[3]{#2}
\let\VANthebibliography\thebibliography
\def\thebibliography{\DeclareRobustCommand{\VAN}[3]{##3}\VANthebibliography}

\usepackage{graphicx}	
\usepackage{amsmath}	

\title[3D effects on WASP-76b]{Modelling the effect of 3D temperature and chemistry on the cross-correlation signal of transiting ultra-hot Jupiters: A study of 5 chemical species on WASP-76b}

\author[Joost P. Wardenier et al.]{
Joost P. Wardenier,$^{1}$\thanks{E-mail: joost.wardenier@physics.ox.ac.uk}
Vivien Parmentier,$^{1,2}$
Michael R. Line,$^{3}$
Elspeth K. H. Lee$^{4}$
\\
$^{1}$Department of Physics (Atmospheric, Oceanic and Planetary Physics), University of Oxford, Oxford, OX1 3PU, UK \\
$^{2}$Université Côte d’Azur, Observatoire de la Côte d’Azur, CNRS, Laboratoire Lagrange, Bd de l’Observatoire, CS 34229, 06304 Nice cedex 4, France \\
$^{3}$School of Earth \& Space Exploration, Arizona State University, Tempe AZ 85287, USA \\
$^{4}$Center for Space and Habitability, University of Bern, Gesellschaftsstrasse 6, CH-3012 Bern, Switzerland
}

\date{Accepted XXX. Received YYY; in original form ZZZ}

\pubyear{2023}

\begin{document}
\label{firstpage}
\pagerange{\pageref{firstpage}--\pageref{lastpage}}
\maketitle

\begin{abstract}
{\color{black}{Ultra-hot Jupiters are perfect targets for transmission spectroscopy. However, their atmospheres feature strong spatial variations in temperature, chemistry, dynamics, cloud coverage, and scale height. This makes transit observations at high spectral resolution challenging to interpret. In this work, we model the cross-correlation signal of five chemical species -- Fe, CO, H$_\text{2}$O, OH, \mbox{and TiO --} on WASP-76b, a benchmark ultra-hot Jupiter. We compute phase-dependent high-resolution transmission spectra of 3D SPARC/MITgcm models. The spectra are obtained with gCMCRT, a 3D Monte-Carlo radiative-transfer code. We find that, on top of atmospheric dynamics, the phase-dependent Doppler shift of the absorption lines in the planetary rest frame is shaped by the combined effect of planetary rotation and the unique 3D spatial distribution of chemical species. For species probing the dayside (e.g., refractories or molecules like CO and OH), the two effects act in tandem, leading to increasing blueshifts with orbital phase. For species that are depleted on the dayside (e.g., H$_\text{2}$O and TiO), the two effects act in an opposite manner, and could lead to increasing redshifts during the transit. This behaviour yields species-dependent offsets from a planet’s expected $K_\text{p}$ value that can be much larger than planetary wind speeds. The offsets are usually negative for refractory species. We provide an analytical formula to estimate the size of a planet’s $K_\text{p}$ offsets, which can serve as a prior for atmospheric retrievals. We conclude that observing the phase-resolved absorption signal of multiple species is key to constraining the 3D thermochemical structure and dynamics of ultra-hot Jupiters.}}

\end{abstract} 

\begin{keywords}
radiative transfer -- methods: numerical -- planets and satellites: atmospheres -- planets and satellites: gaseous planets
\end{keywords}



\section{Introduction}
\label{sec:intro}

Ultra-hot Jupiters are an extreme class of exoplanet with equilibrium temperatures greater than $\sim$2000 K (\citealt{Bell2018,Arcangeli2018,Parmentier2018}). They offer a unique opportunity to study atmospheric physics and chemistry under conditions that do not prevail on any of the planets in our own Solar System. To date, the formation history of ultra-hot Jupiters is largely unknown, but constraining the elemental abundance ratios of their atmospheres can shed light on their origins, accretion mechanisms, and their migration through the protoplanetary disk (\citealt{Oberg2011,Mordasini2016,Espinoza2017,Line2021,Lothringer2021,Schneider2021a,Schneider2021b,Molliere2022,Hobbs2022}). Ultra-hot Jupiters are ideal targets for atmospheric characterisation in transmission, thanks to their extended atmospheres, short orbital periods (1$-$2 days), and simple chemical inventory. However, one aspect that complicates the interpretation of their spectra is their inherent ``3D-ness'' (\citealt{Flowers2019,Caldas,Lacy2020,Pluriel2020,Pluriel2022,Wardenier2021,Falco2022,Savel2022,Gandhi2022,Pluriel2023,Beltz2023}). Ultra-hot Jupiters are tidally locked, which means that they have a permanent dayside and a permanent nightside with very different temperature structures and chemical compositions. On the hot, puffy dayside refractories and alkalis such as Fe, Mg, Ca, Ba, K, and Na exist in their atomic or ionised form, while molecules such as H$_2$, H$_2$O, and TiO get thermally dissociated. On the nightside, the temperature is much lower, allowing for cloud formation to occur (\citealt{Wakeford2017,Helling2019,Helling2021,Parmentier2021,Komacek2022}). The large day-night contrast results in steep thermochemical gradients and scale-height variations across the terminator region of the atmosphere, which is probed by transmission spectroscopy. Additionally, the day-night contrast drives fast winds in the order of \mbox{1$-$10 km/s} (\citealt{Tan2019}). The wind profile of ultra-hot Jupiters can be decomposed into two contributions: a day-to-night flow that carries material from the dayside to the nightside of the planet, and (depending on the drag conditions in the atmosphere) a superrotating jet around the equator (\citealt{Showman2008,Komacek2016,Komacek2017,Hammond2021}).

Arguably the best technique for studying ultra-hot Jupiters is ground-based high-resolution spectroscopy (HRS -- \citealt{Snellen2010, Brogi2012, Birkby2018}). Thanks to its ability to \emph{resolve individual spectral lines} and perform local measurements, HRS can shed light on atmospheric physics that is not accessible to low-resolution (i.e., HST and JWST) observations. As a planet orbits its star, its radial velocity changes and its spectral lines are periodically Doppler-shifted. This allows for the planet signal to be isolated from stellar and telluric contributions. Over the past few years, planets such as WASP-33b (e.g., \citealt{Cont2021,Nugroho2021,vanSluijs2022}), WASP-76b (e.g., \citealt{Seidel2019,Ehrenreich2020,Landman2021}), WASP-121b (e.g., \citealt{Gibson2020,Borsa2021,Merritt2021}), KELT-9b (e.g., \citealt{Hoeijmakers2018,Pino2020,Kasper2021}), and KELT-20b (e.g., \citealt{Casasayas-Barris2018,Nugroho2020,Yan2022}) have been targeted by a large number of HRS observations, both in the optical and the infrared. These have enabled the detection of a plethora of chemical species\footnote{See Table 1  in \citet{Guillot2022} for a relatively recent overview of detected species in the atmospheres of gas giants.}, as well as wind-speed measurements (e.g., \citealt{Casasayas-Barris2018,Bourrier2020,Seidel2021}). Additionally, for various ultra-hot Jupiters, HRS observations revealed evidence for hot-spot shifts and thermal inversions on the dayside (e.g., \citealt{Pino2020,Kasper2021,Herman2022,vanSluijs2022}), cloud formation on the nightside, and asymmetries between the morning and evening limbs of the planet (\citealt{Ehrenreich2020,Kesseli2021,Borsa2021,Kesseli2022,Sanchez-Lopez2022,Pelletier2023}).

At high resolution, the ``3D-ness'' of ultra-hot Jupiters causes the absorption lines in their transmission spectrum to be shifted, broadened, and distorted (e.g., \citealt{Seidel2021,Wardenier2021,Keles2021}). This is because stellar light rays encounter different pressures, temperatures, abundances, and line-of-sight velocities 
as they pass through the atmosphere. A few lines are strong enough to be seen directly, but the vast majority of the planet spectrum lies buried in stellar photon noise. 
{\color{black}{One way to detect the planet signal is to cross-correlate the spectrum with a template model and \emph{combine} the strengths of all the absorption lines (typically associated with a single chemical species).}} 
This results in a cross-correlation function (CCF -- \citealt{Snellen2010,Snellen2015,Rodler2013,Hoeijmakers2019}), which is a measure for the similarity between the planet spectrum and the template as a function of radial velocity (i.e., Doppler shift). The total Doppler shift of the planet spectrum is induced by the systemic velocity $V_\text{sys}$ of the star, the orbital velocity $K_\text{p}$ of the planet, its rotation, and its atmospheric dynamics. However, since the ($K_\text{p}$, $V_\text{sys}$) values of a planet are known, it is possible to transform the CCF to a \emph{planetary rest frame}, in which the only Doppler contributions are from rotation and dynamics. These ``anomalous'' Doppler shifts contain information about the 3D nature of the planet.

Because ultra-hot Jupiters are tidally locked, they rotate by \mbox{25--30} degrees during their transit (\citealt{Wardenier2022}), assuming an edge-on orbit. This means that the transmission spectrum may probe different parts of the atmosphere at different orbital phases. At the start of the transit, the leading limb (or \emph{morning limb}) is largely comprised of dayside atmosphere, while the trailing limb (or \emph{evening limb}) mainly covers the nightside. Then, as the transit progresses, the dayside rotates into view on the trailing limb, and the nightside rotates into view on the leading limb (\citealt{Ehrenreich2020,Wardenier2021}). Because the terminator regions of ultra-hot Jupiters are characterised by extreme spatial variations in temperature, chemistry, dynamics, and scale height, the rest-frame CCF can be expected to undergo substantial changes over the course of the transit.

From an observational standpoint, however, ``phase-resolving'' the absorption signal of a species in a transiting exoplanet atmosphere is a challenge. To our knowledge, this has only been attempted for \mbox{WASP-76b} (\citealt{Ehrenreich2020,Kesseli2021,Kesseli2022,Sanchez-Lopez2022,Pelletier2023}), \mbox{WASP-121b} (\citealt{Bourrier2020,Borsa2021}), and \mbox{KELT-20b} (\citealt{Rainer2021}). 

For WASP-76b and WASP-121b, the CCFs of neutral iron (Fe or Fe \textsc{i}) show an increasing blueshift during the transit, with the peak position moving from about 0 km/s at ingress to about $-$10 km/s at egress (\citealt{Ehrenreich2020,Borsa2021}). {\color{black}{In the case of WASP-76b, the absorption trail features a ``kink'' around mid-transit in the CCF map (e.g., Fig. 1 in \citealt{Wardenier2021})}}. Multiple mechanisms were suggested for this behaviour, including iron condensation on the leading limb of the planet (\citealt{Ehrenreich2020, Wardenier2021}), a scale-height (temperature) difference between both limbs (\citealt{Wardenier2021}), the presence of optically-thick clouds on the leading limb (\citealt{Savel2022}), or a combination of these effects. More recently, \citet{Beltz2023} proposed that the planet's (spatially varying) magnetic field can also play a role.

Using the VLT/ESPRESSO dataset from \citet{Ehrenreich2020}, \citet{Kesseli2022} went on to study the phase-dependent behaviour of a large number of other species in WASP-76b besides iron, namely 
H, Li, Na, Mg, K, Ca \textsc{ii}, V, Cr, Mn, Co, Ni, and Sr \textsc{ii}. {\color{black}{They found that the CCFs of \emph{all species} except atomic hydrogen and lithium were \emph{more blueshifted} in the final quarter of the transit compared to the first quarter. Recent GEMINI-N/MAROOX-X observations of WASP-76b by \citet{Pelletier2023} confirmed these trends. Moreover, \citet{Pelletier2023} reported that the vast majority refractories and alkalis -- \emph{species expected to be abundant on the dayside} -- give rise to absorption trails with the same ``kink'' feature as iron\footnote{Ionised calcium (Ca \textsc{ii}) is an exception, as its absorption originates from higher regions in the atmosphere which are likely subject to atmospheric escape.}. Based on these observations, the authors suggested that the iron signal of WASP-76b is shaped by a global mechanism that also affects other species in the optical, rather than condensation alone.}}

In the infrared, \citet{Sanchez-Lopez2022} used CARMENES data to measure the H$_2$O and the HCN signals of WASP-76b. They also found substantial differences between the first and second half of the transit, both in terms of Doppler shift and CCF strength. 

\begin{table*}
\centering
\caption{Overview of the four WASP-76b scenarios considered in this work, based on outputs of the SPARC/MITgcm.}
\makebox[\textwidth][c]{
\begin{tabular}{l|c|c}
\hline
\textbf{Model} & \textbf{Drag timescale} & \textbf{Description} \\ \hline
Nominal model & $10^5$ s & Nominal GCM output without post-hoc additions or modifications.\\
Cold morning limb & $10^5$ s & Nominal GCM output with modified thermochemical structure on the morning limb, following \citet{Wardenier2021}. \\
Optically thick clouds & $10^5$ s & Nominal GCM output with optically thick Al$_2$O$_3$ cloud added, following \citet{Savel2022}.\\
No TiO/VO & $\rightarrow \infty$ & Atmosphere without TiO and VO opacities \emph{during} the GCM calculations; no post-hoc additions or modifications.\\
\hline
\end{tabular}
}
\label{tab:gcm_models}
\end{table*}

{\color{black}{Transit observations resolved with orbital phase are a powerful means to perform local measurements in an exoplanet atmosphere and thus obtain information about its ``3D-ness''.}} For example, by dividing the VLT/ESPRESSO dataset from \citet{Ehrenreich2020} into two halves, \citet{Gandhi2022} were able to retrieve the temperature profile and iron abundance of WASP-76b at \emph{four} different longitudes. Furthermore, they separately constrained the wind speeds on the trailing and leading limb of the planet. Performing similar retrieval studies in the infrared would be valuable for two reasons. Firstly, they allow to get a better handle on the planet's ``3D-ness'', as different species probe different atmospheric regions. Secondly, measuring the abundances of molecules such as CO, H$_2$O, and OH allows to compute refractory-to-volatile ratios (e.g., Fe/O), which are important in the context of planet formation (\citealt{Lothringer2021,Feinstein2023}). However, the fact that the planet is 3D will make it more difficult to make these inferences, {\color{black}{as abundances vary spatially}}. Therefore, we need 3D forward models to understand \emph{how} 3D effects manifest in high-resolution spectra and \emph{how} to best parameterise these effects in 1D or pseudo-2D models used in retrievals. Also, we require 3D forward models to understand what we can really learn from multi-species observations.  

The aim of this work is to further explore the connection between the ``3D-ness'' ultra-hot Jupiters and their CCF signals in transmission. To this end, we  build on earlier modelling work described in \citet{Wardenier2021}. We use a 3D Monte-Carlo radiative transfer framework to simulate phase-dependent transmission spectra for different atmospheric scenarios of WASP-76b, based on outputs of a global circulation model (GCM). We then compute the CCF signals and $K_\text{p}$--$V_\text{sys}$ maps for five different chemical species: Fe and TiO in the optical, and CO, H$_2$O, and OH in the infrared. The motivation for considering these species is that they all have \emph{distinct 3D spatial distributions} across the planet. Therefore, the absorption lines associated with these species will probe different regions of the atmosphere, each with their own properties. Furthermore, the behaviour we identify for a certain species will be representative of other atoms and molecules with the same spatial distribution. For example, the signals we simulate for iron will be a good proxy for the signals of other refractories too. 

The structure of this manuscript is as follows. In Section \ref{sec:methods} we describe our WASP-76b models, our radiative-transfer framework, and methods for computing CCF signals, $K_\text{p}$--$V_\text{sys}$ maps and absorption regions. In Section \ref{sec:results}, we present, discuss and interpret our results. Finally, Section \ref{sec:conclusion} provides a conclusion.

\section{Methods}
\label{sec:methods}

\subsection{Model atmospheres}
\label{subsec:model_atmospheres}

\subsubsection{General overview}

In this work, we consider four different 3D models of the atmosphere of WASP-76b, based on outputs of the SPARC/MITgcm  global circulation model (\citealt{Showman2009}). For the setup of the GCM simulations, we refer the reader to \citet{Parmentier2018} and \citet{Wardenier2021}. {\color{black}{All models assume solar values for metallicity and C/O ratio. We compute the abundances through chemical equilibrium, such that the number fraction of a species in a given atmospheric cell only depends on the local pressure and temperature.}} In addition, the GCM accounts for condensation through ``rainout'', whereby a certain fraction of a species (e.g., Fe or Mg) is removed from a cell when the local temperature lies below the (pressure-dependent) condensation temperature of a condensate containing that species (e.g., \citealt{Visscher2010}). The process is called rainout, because it assumes that condensates instantly settle to a deeper layer where they do not impact the radiation balance of the atmosphere. 


The four models are summarised in Table \ref{tab:gcm_models}. Our nominal model is the same as the weak-drag model from \citet{Wardenier2021}. It has a drag timescale $\tau_\text{drag} = 10^5$ s, which has been found to provide a better match to the WASP-76b observations than a drag-free atmosphere (\citealt{May2021,Wardenier2021,Savel2022,Gandhi2022}). The drag timescale represents the typical time it takes for an air parcel to lose a significant fraction of its kinetic energy. It encapsulates a number of different processes, such as turbulent mixing (\citealt{Li2010}), Lorentz-force braking of winds of charged particles due to the planet's magnetic field (\citealt{Perna2010}), and Ohmic dissipation (\citealt{Perna2010a}). As a result of drag forces, the equatorial jet of the planet is suppressed, such that the atmospheric dynamics are dominated by a day-to-night flow. 

The second model we consider is the cold-morning-limb model from \citet{Wardenier2021} (see ``modification 2'' in their Fig. 12), in which we artificially reduce the temperature of the leading limb. Consequently, the atmosphere features a strong thermal asymmetry between the (hotter) trailing limb and (cooler) leading limb. In \citet{Wardenier2021}, we demonstrated that this model is able to reproduce the shape of the iron signal of WASP-76b (\citealt{Ehrenreich2020,Kesseli2021}), as opposed to an atmosphere without an east-west asymmetry. In the cold-morning-limb model, the absorption lines undergo an increasing blueshift during the first half of the transit, but they retain a constant Doppler shift of about \mbox{$-$8 km/s} during the second half.

Our third model is an optically-thick-clouds model à la \citet{Savel2022}, who reported that the iron signal of \mbox{WASP-76b} can also result from an atmosphere with optically thick clouds. In one of their best-fitting models, they assume the presence of an optically thick cloud deck extending at most 10 scale heights ($\sim$4.3 dex in pressure) above the intersection between the local temperature profile and the Al$_2$O$_3$ condensation curve. The vertical extent of the cloud is less than 10 scale heights in case the temperature profile and the condensation curve intersect again at some lower pressure. We add an optically thick cloud deck to our GCM output in exactly the same way (clouds are added post-hoc, so the temperature structure in the GCM is calculated \emph{without} clouds). The rationale for selecting Al$_2$O$_3$ is that this is the cloud species with the highest condensation temperature. Hence, it will have the most drastic impact on the planet's transmission spectrum, as it can exist in hotter regions compared to other cloud species. Because cloud physics is complicated (e.g., \citealt{Helling2006,Gao2018a,Gao2018b}), this modelling approach is very much a simplification of reality. However, the model forms a good \emph{limiting case} -- it allows to assess the strongest impact that clouds can possibly have on the CCF signal.

Our final model is the atmosphere without TiO and VO from \citet{Wardenier2021}. It represents a scenario in which TiO and VO are cold-trapped due to condensation (\citealt{Spiegel2009,Parmentier2013,Beatty2017}). To emulate the effects of cold-trapping, the opacities of TiO and VO are set to zero during the GCM calculations. Because these molecules are important short-wave absorbers, their absence will change both the dynamics and the temperature structure of the atmosphere. As shown in \citet{Wardenier2021}, the no-TiO/VO model naturally has a large temperature asymmetry between its trailing and leading limb, owing to a strong hotspot shift on the dayside that extends to relatively low pressures. Furthermore, our no-TiO/VO model is drag-free ($\tau_\text{drag} \rightarrow \infty$), so it features an equatorial jet. The model provides a good test to assess which observational features are robust against a variety of different modelling assumptions. 

\subsubsection{Mapping pressures onto altitudes}

As described in \citet{Wardenier2021}, the GCM uses pressure as a vertical coordinate. However, to compute the transmission spectrum of the planet, the atmosphere must be defined on an altitude grid. Thus, before we can feed the models into the radiative-transfer framework, we need to perform the mapping $P \rightarrow z$ in every atmospheric column, with $P$ the pressure and $z$ the altitude coordinate. To this end, we follow the approach from \citet{Wardenier2021} (see their Section 3.2), whereby we assume that the atmosphere is an ideal gas in hydrostatic equilibrium. For every atmospheric cell $i$, we compute the scale height as follows:

\begin{equation} \label{eq:scale_height}
    H_i = \frac{k_B T_i}{\mu_i g_i},
\end{equation}

\noindent with $k_B$ the Boltzmann constant, $T_i$ the cell's temperature, $\mu_i$ its mean-molecular weight, and $g_i$ its gravity.

One important improvement we make compared to \citet{Wardenier2021} is that we also account for mean-molecular weight variations across the atmosphere when computing $H_i$, in addition to temperature and gravity variations. On most of the dayside, the mean-molecular weight is significantly lower than on the nightside due to hydrogen (H$_2$) dissociation -- lowering its value from $\mu$ $\approx$ 2.33 $m_\textsc{h}$ to $\mu$ $\approx$ 1.27 $m_\textsc{h}$ (with $m_\textsc{h}$ the mass of a hydrogen atom). As a result, the scale-height difference between the dayside and the nightside of the models is even larger than suggested in \citet{Wardenier2021}. We verified, however, that (not) accounting for thermal dissociation in the $P \rightarrow z$ mapping does not drastically alter the shape of the final CCF signals (see Appendix \ref{ap:A}), so the results from \citet{Wardenier2021} remain valid.

\begin{figure}
\centering
\vspace{-25pt}
\includegraphics[width=0.48\textwidth]{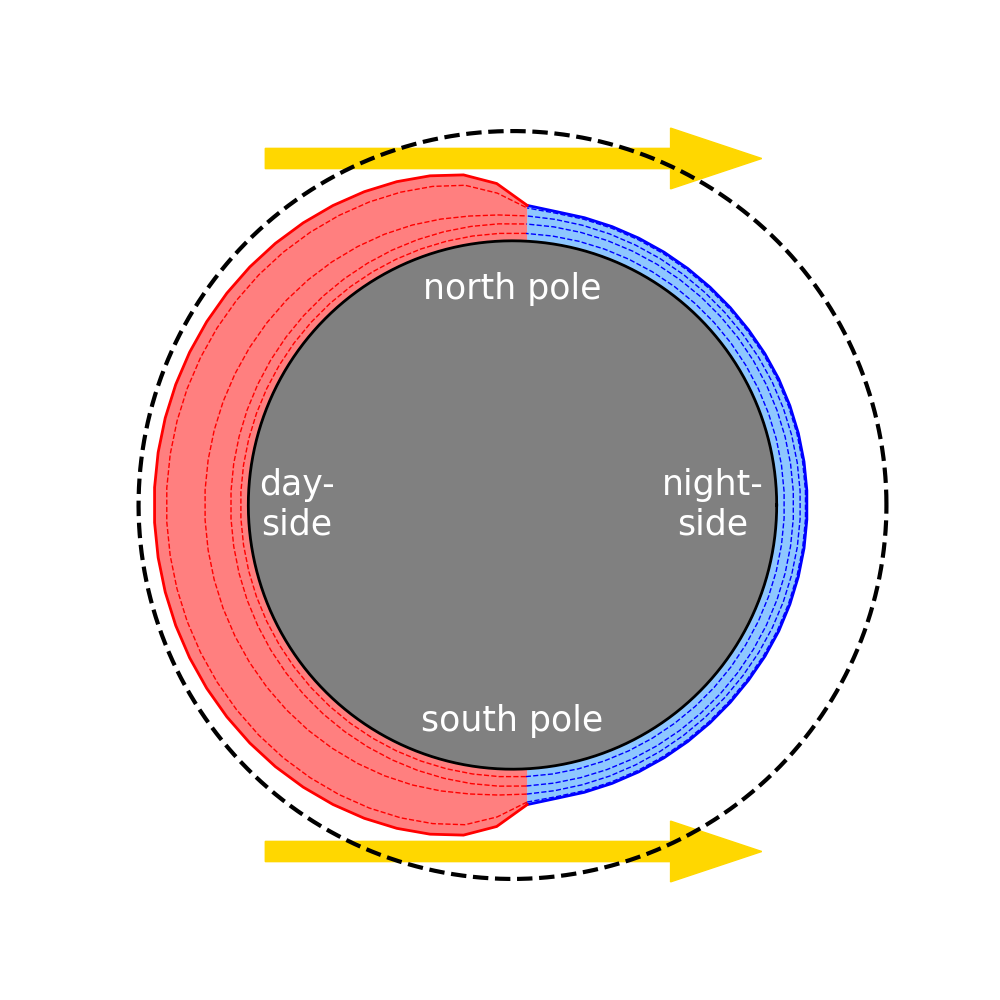}
\vspace{-30pt}
\caption{A cross-section of the nominal model of WASP-76b, with the interior (grey region), the dayside atmosphere (red region), and the nightside atmosphere (blue region) plotted to scale. The dayside is more ``puffy'' than the nightside due to its higher temperature and lower mean-molecular weight. Dashed lines indicate isobars with pressures $10^1$, $10^{-1}$, $10^{-3}$ and $10^{-5}$ bar, respectively. The solid line marks the upper GCM boundary at 2 $\mu$bar. When mapping the GCM output onto the altitude grid, we extrapolate the atmosphere to the same altitude everywhere (indicated by the black dashed circle). This means that the nightside extends to much lower (negligible) pressures than the dayside. Yellow arrows indicate light rays pointing towards the observer.}
\label{fig:cross_cut_gcm}
\end{figure}

Fig. \ref{fig:cross_cut_gcm} shows a \emph{to-scale} plot of the nominal model mapped onto its altitude grid. The bottom of the atmosphere, with \mbox{$P$ = 200 bar} and \mbox{$g$ = 7.6 m/s}, is situated at a radius $R = 1.85$ $R_\text{jup}$. At the substellar point (on the dayside), the 10-$\mu$bar isobar lies at $R = 2.44$ $R_\text{jup}$. At the antistellar point (on the nightside), it lies at $R = 2.06$ $R_\text{jup}$. To prevent absorption lines from being ``truncated'' by the model boundaries in the radiative transfer, we extrapolate the entire atmosphere to a radius $R$ = 2.64 $R_\text{jup}$ (black dashes in Fig. \ref{fig:cross_cut_gcm}), assuming that temperatures, abundances, and wind speeds remain constant above the upper GCM boundary of 2 $\mu$bar. Because the nightside has a much smaller scale height, it is extrapolated to trivially low pressures where the absorption is zero.

\subsection{Radiative transfer}
\label{subsec:radiative_transfer}

\subsubsection{Monte-Carlo radiative transfer with gCMCRT}

To compute transmission spectra associated with the 3D model atmospheres, we use gCMCRT\footnote{gCMCRT is publicly available from \href{https://github.com/ELeeAstro/gCMCRT}{\texttt{github.com/ELeeAstro/gCMCRT}} } (\citealt{Lee2022}). gCMCRT is an updated, GPU-compatible version of Monte-Carlo radiative transfer code from \citet{Lee2017,Lee2019}. In \citet{Wardenier2021}, we adapted the framework for high-resolution purposes. The main advantage of gCMCRT is that it fully exploits the architecture of a GPU, which comprises hundreds to thousands of individual cores (processing units). Hence, a large number of photon packets can be simulated in parallel, making gCMCRT a lot faster than its predecessor. In \citet{Wardenier2021}, we had to restrict our simulations to $\sim$10,000 wavelength points for computational reasons, but with gCMCRT we can efficiently model high-resolution spectra across the full bandwidth of instruments like VLT/ESPRESSO and Gemini-S/IGRINS.  

For each of the four WASP-76b models, we simulate the orbit over an angle of 31.3 degrees, covering the transit as well as ingress and egress. We compute 25 transmission spectra, equidistant in orbital phase. Furthermore, we assume an edge-on orbit, a semi-major axis of 0.033 AU, a stellar radius of 1.73 $R_{\text{sun}}$, and an orbital period of 1.81 days -- commensurate with the parameters of the WASP-76 system (\citealt{West2016}). We ignore effects of limb darkening as these were reported have a negligible impact on the Doppler shifts obtained from cross-correlation (\citealt{Savel2022}).


As discussed in \citet{Wardenier2021}, Monte-Carlo radiative transfer is a stochastic technique. To compute a transmission spectrum, we initialise $n$ photon packets with a random impact parameter and impact angle at each wavelength. During ingress and egress we only illuminate the part of the limb that is blocking the star. The spectrum converges to the true solution in the limit $n \rightarrow \infty$ (we use $n = 10^5$ in this work). For each photon packet, we compute the optical depth $\tau$ along the line of sight, whereby we Doppler-shift the opacities in each atmospheric cell according to the local line-of-sight velocity $v_\textsc{los}$ that results from winds and planetary rotation (see Fig. \ref{fig:winds}). We refer to Section 3.3 in \citet{Wardenier2021} for the relevant equations. Because we account for scattering through absorption cross-sections (a treatment justified in transmission as scattering causes photons to \emph{depart} from the line of sight and not contribute to the flux), we effectively use gCMCRT as a randomised-transit-chord algorithm. The propagation direction of the photon packets does not change after their initialisation. 

Once the optical depth associated with the photon packets has been computed, the ``transit area'' $A_{\text{p}}(\lambda)$ of the planet can be found from (\citealt{Wardenier2021,Wardenier2022})

\begin{equation} \label{eq:def_transit_area}
    A_{\text{p}}(\lambda) = A_{\text{0}} + A_{\text{annu}} \big\langle 1 - e^{-\tau} \big\rangle \big\rvert_{\lambda},
\end{equation}

\noindent with $A_{\text{0}}$ the projected area of the planetary interior and $A_{\text{annu}}$ the area of the atmospheric annulus (extending from the bottom to the top of the model atmosphere). The angle brackets imply an average over all photon packets with wavelength $\lambda$. During ingress and egress, we scale down the value of $A_\text{p}$ with the fractional overlap ($< 1$) between the stellar and the planetary disk to obtain the correct transit depth.  

As in \citet{Wardenier2021}, we also compute spectra associated with individual sectors on the limb (see their Fig. 3): the trailing equator, the trailing pole(s), the leading pole(s), and the leading equator. The trailing \emph{(leading)} equator is the limb region between $-45^\circ$ and $+45^\circ$ latitude that is last \emph{(first)} to appear in front of the star during ingress. The trailing \emph{(leading)} poles are the regions between $-90^\circ$ and $-45^\circ$, and $+45^\circ$ and $+90^\circ$ that are last \emph{(first)} to appear in front of the star during ingress. All sectors span a quarter of the limb, but as shown in Fig. 3 in \citet{Wardenier2021}, the poles are disjoint. For a tidally locked planet, the trailing regions rotate towards the observer, while the leading regions rotate towards the star (away from the observer). To compute spectra for each sector, we also use equation \ref{eq:def_transit_area}, but we only perform the average over the photon packets impinging on that sector.  

\subsubsection{Modelling spectra in the optical (Fe and TiO signals)}
\label{subsec:optical_spectra}

In the optical, we model the transit of the four WASP-76b models across the full ESPRESSO wavelength range (0.38--0.79 $\mu$m) at a spectral resolution $R$ = 300,000 {\color{black}{($>$$2\times$ the ESPRESSO resolution)}}. This results in a total of $\sim$220,000 wavelength points\footnote{{\color{black}{With 10$^{5}$ photon packets per wavelength, this means that the total number of photon packets simulated across the spectrum is of the order 10$^{10}$.}}}. For memory-related reasons, we split the computation in two batches of $\sim$110,000 wavelength points and we stitch the spectra together at the end. Since we read all opacity data at once at the start of the simulation, the GPU memory needs to hold the full 3D opacity structure of the atmosphere at \emph{each} wavelength.

In the radiative transfer, we include (continuum) opacities associated with H$_2$, He, and H scattering, collision-induced absorption (CIA) by H$_2$-H$_2$ and H$_2$-He, and bound-free and free-free transitions of H$^-$. References to these opacities can be found in Table 2 in \citet{Lee2022b}. Also, we consider the following line species: Fe, Fe \textsc{ii}, K, Na, Ti, Mn, Mg, Cr, Ca \textsc{ii}, TiO, VO, H$_2$O, and OH. Atomic opacities are taken from the \citet{Kurucz1995} database and we apply pressure broadening using a code based on \citet{Kurucz1981}. In \citet{Wardenier2021}, the atomic opacities were generated with \texttt{HELIOS-K} (\citealt{Grimm2021}) and no pressure broadening was applied. Furthermore, \texttt{HELIOS-K} imposed a line-wing cut-off, as opposed to our current treatment. The opacities of TiO and VO are from the {\tt EXOPLINES} database (\citealt{Gharib-Nezhad2021-ZENODO}), and were generated by \citet{Gharib-Nezhad2021-APJSpaper} using the TOTO (\citealt{Mckemmish2019}) and the VOMYT (\citealt{McKemmish2016}) line lists. For H$_2$O, we use the POKAZATEL line list (\citealt{Polyansky2018}). Finally, the OH opacities are taken from \texttt{HITEMP} (\citealt{Rothman2010}).

Compared to \citet{Wardenier2021}, we thus make a  total of four changes to the radiative transfer. Firstly, we use iron line lists with pressure broadening and no line-wing cut-off, and we use opacities for a larger number of species. Secondly, we account for variations in mean-molecular weight when evaluating the scale height (see Section \ref{subsec:model_atmospheres}). Thirdly, we reduce the spectral resolution from 500,000 to 300,000. Finally, we consider the full ESPRESSO wavelength range instead of a small set of $\sim$10,000 wavelength points. Fig. \ref{fig:appendix_plot} in Appendix \ref{ap:A} depicts the effect that each of these changes has on the iron signal of the cold-morning-limb model originally presented in \citet{Wardenier2021}. The figure shows that the ``new'' iron opacities and the new resolution do not significantly impact the CCF map. As expected, the new scale heights and the new wavelength range lead to the biggest changes, but the overall trends in the CCF map remain the same. 

\subsubsection{Modelling spectra in the infrared (CO, H$_\text{2}$O, and OH signals)}

In the infrared, we model the transit of the four WASP-76b models across the full IGRINS wavelength range (1.43--2.42 $\mu$m) at \mbox{$R$ = 135,000} ($\sim$3$\times$ the IGRINS resolution). This leads to $\sim$71,000 wavelength points. We can afford a lower resolution here as the absorption features of the relevant molecules tend to be intrinsically broader than in the optical, so they can still be resolved at a lower resolution. We performed a comparison similar to Fig. \ref{fig:appendix_plot} to verify that the Doppler shifts obtained from cross-correlation remain the same (within 0.5 km/s) at higher spectral resolutions.

To compute the infrared spectra, we consider the same continuum opacities as in the optical. Additionally, we use the line species CO (\citealt{Li2015}), H$_2$O (\citealt{Polyansky2018}), OH (\citealt{Rothman2010}), CH$_4$ (\citealt{Hargreaves2020}), CO$_2$ (\citealt{Rothman2010}), HCN (\citealt{Barber2014}), and NH$_3$ (\citealt{Coles2019}).



\subsection{Computing observables}
\label{subsec:computing_observables}

\begin{figure*}
\vspace{-10pt} 
\makebox[\textwidth][c]{\hspace{-35pt}\includegraphics[width=1.15\textwidth]{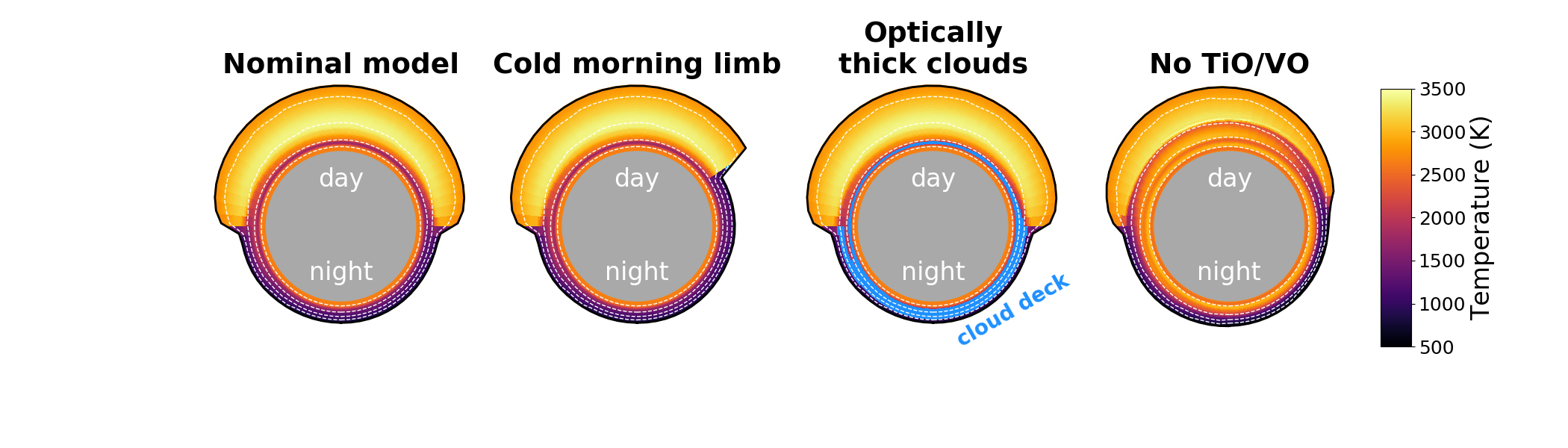}}
\vspace{-45pt}
\caption{The temperature in the equatorial plane of the four models considered in this work (see Table \ref{tab:gcm_models}). Note that the relative size of the atmosphere has been exaggerated for visualisation purposes -- see Fig. \ref{fig:cross_cut_gcm} for the correct scale. White dashes indicate isobars with pressures $10^1$, $10^{-1}$, $10^{-3}$ and $10^{-5}$ bar, respectively. The solid black line marks the upper GCM boundary at 2 $\mu$bar. The blue patch in the optically-thick-clouds model shows the extent of the Al$_2$O$_3$ clouds, which are modelled according to the description in \citet{Savel2022}.}
\label{fig:eq_temps}
\end{figure*}

\subsubsection{CCF maps}

For each transit, we cross-correlate all 25 spectra with a template -- see Section 3.6 in \citet{Wardenier2021} for relevant equations. This gives rise to a CCF map with Doppler shift (radial velocity, or RV) as a horizontal coordinate and orbital phase as a vertical coordinate. We compute CCF maps for Fe and TiO (based on the optical spectra), as well as for CO, H$_2$O, and OH (based on the infrared spectra). To generate the template for a species $X$, we compute the mid-transit spectrum of the nominal model \emph{without} Doppler shifts, whereby we only include the opacities of the continuum and $X$. Before we perform the cross-correlation, we subtract the continuum from both the templates and the spectra. We do this by splitting a spectrum into bins of 1000 wavelength points and fitting a low-order polynomial to the minima of these bins. We then subtract the polynomial from the spectrum to obtain a ``flat'' baseline. This procedure mimics the steps taken in the analysis of real high-resolution data. 

We also compute CCF maps associated with the four limb sectors. As demonstrated in \citet{Wardenier2021}, the CCF map of the full limb can be interpreted as the sum of the CCF maps of the individual sectors, thanks to the linearity of the cross-correlation. The benefit of this approach is that it allows to link certain features of the CCF map to specific atmospheric regions. 

\subsubsection{$K_\text{p}$--$V_\text{sys}$ maps}

In most high-resolution datasets, the CCF values associated with individual integrations must be ``stacked'' across the whole transit to get a strong enough planet detection. A common way to do this is by constructing a \mbox{$K_\text{p}$--$V_\text{sys}$} map (e.g., \citealt{Brogi2012,Wardenier2021}). The signal emerging in the \mbox{$K_\text{p}$--$V_\text{sys}$} map can be seen as a time average, because it is a sum over all orbital phases.

Once the CCF map of a certain species is computed, we obtain the corresponding $K_\text{p}$--$V_\text{sys}$ map by integrating the CCF values along a curve of the form

\begin{equation} \label{eq:kpvsys_curve}
    v(\phi) = V_\text{sys} + K_\text{p} \sin(\phi),
\end{equation}

\noindent with $v(\phi)$ the radial velocity at phase angle $\phi \in [-15.7^\circ, +15.7^\circ]$, $V_\text{sys}$ the systemic velocity, and $K_\text{p}$ the orbital velocity. In other words:

\begin{equation} \label{eq:kpvsys_sum}
    \text{SNR}(K_\text{p}, V_\text{sys}) = \frac{1}{\xi} \sum_i^{N_\phi} \text{CCF} \Big(\phi_i, v(\phi_i) \Big).
\end{equation}

\noindent In this equation, SNR is the value of the $K_\text{p}$--$V_\text{sys}$ map at $(K_\text{p},V_\text{sys})$, {\color{black}{$\xi$ is a scaling factor}}, and $N_{\phi}$ the number of simulated transit spectra. For each orbital phase, we obtain the CCF value at $v(\phi_i)$ by linearly interpolating between the two values at the nearest radial velocities in the CCF map.

\subsection{Computing absorption regions}
\label{subsec:computing_absorptionregions}

Following the approach from \citet{Wardenier2022}, we also compute absorption regions for each of the atmospheric models (see their Section 3.2.2). The information needed to infer these regions is a byproduct of the radiative transfer. 

The idea is that the spectrum does not contain any information about parts of the atmosphere where all the light is absorbed ($e^{-\tau}$$\sim$0) or where all the light is transmitted ($e^{-\tau}$$\sim$1). Instead, the observation probes the region where the \emph{transition} from optically thick to optically thin occurs. Hence, given a wavelength $\lambda$, we define the absorption region as being spanned by all transit chords that satisfy $\beta < e^{-\tau} < 1 - \beta$. In \citet{Wardenier2022}, we opted for $\beta = 0.1$ and $\beta = 0.01$, and we named the corresponding regions the 10--90$\%$ and the 1--99$\%$ absorption regions, respectively. 

The condition $\beta < e^{-\tau} < 1 - \beta$ only constrains the extent of the absorption regions in the altitude direction. However, to obtain a region that is finite \emph{along} the line of sight as well, we only select the central part of the transit chords where the total optical depth increases from $\beta\tau$ to $(1-\beta)\tau$\footnote{As motivated in \citet{Wardenier2022}, this definition ensures that an absorption region is symmetric about the limb plane in the limit of a uniform 1D atmosphere.}. These two conditions allow us to infer the approximate regions that are probed by the transmission spectrum at a certain wavelength. Because we ``truncate'' the transit chords along the line of sight, the extent of the absorption regions is also independent of the (arbitrary) upper model boundary.

\begin{figure*}
\vspace{-10pt} 
\makebox[\textwidth][c]{\hspace{-5pt}\includegraphics[width=1.1\textwidth]{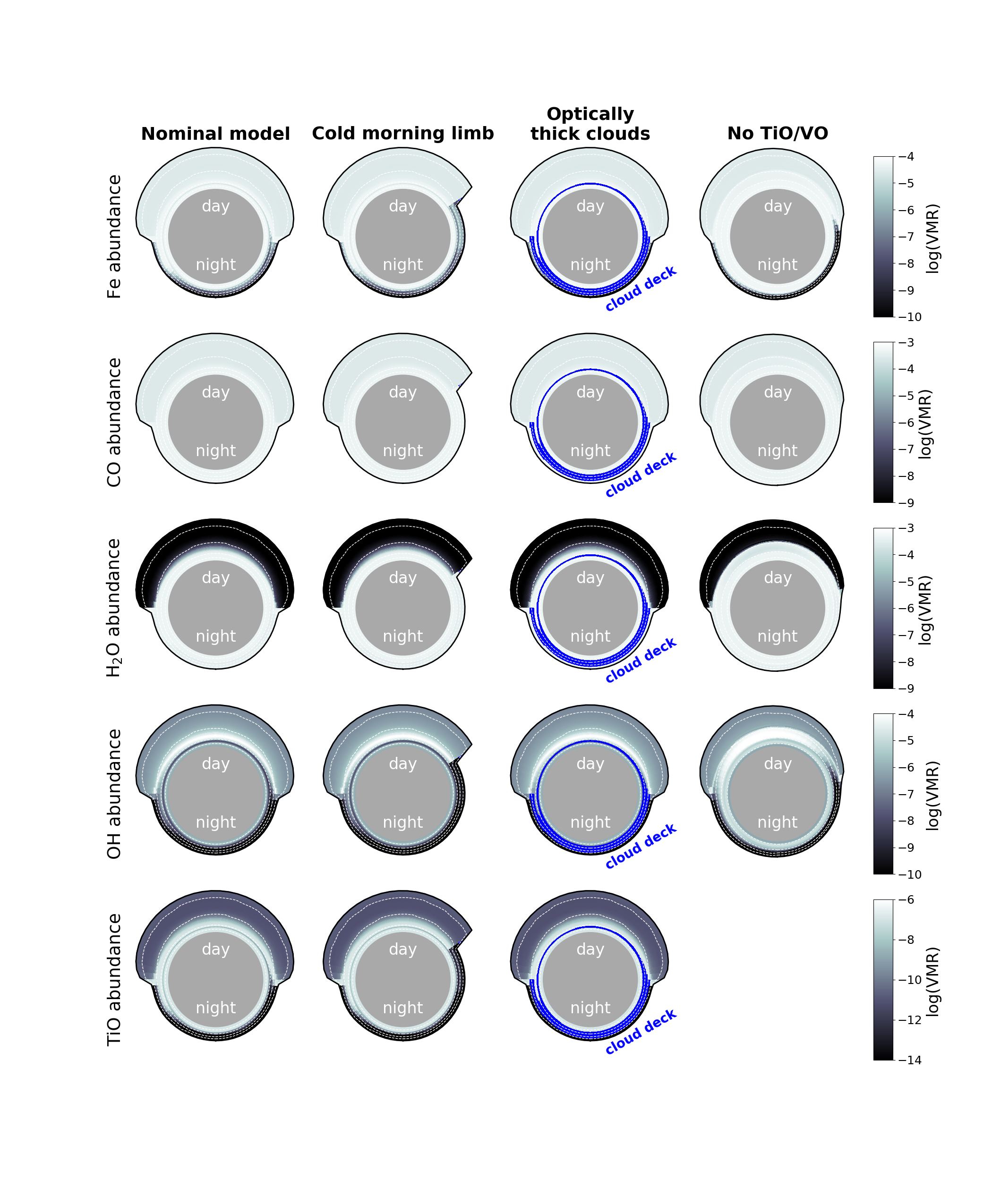}}
\vspace{-80pt}
\caption{The abundances of Fe, CO, H$_2$O, OH, and TiO (one species per row) in the equatorial plane of the four models considered in this work (see Table \ref{tab:gcm_models}). Note that the relative size of the atmosphere has been exaggerated for visualisation purposes. White dashes indicate isobars with pressures $10^1$, $10^{-1}$, $10^{-3}$ and $10^{-5}$ bar, respectively (the planet's transmission spectrum probes pressures below $10^{-1}$ bar). The solid black line marks the upper GCM boundary at 2 $\mu$bar. The dark blue patch in the optically-thick-clouds model shows the extent of the Al$_2$O$_3$ clouds, which are modelled according to the description in \citet{Savel2022}.}
\label{fig:eq_abunds}
\end{figure*}

\begin{figure*}
\centering
\vspace{-20pt} 
\makebox[\textwidth][c]{\hspace{-5pt}\includegraphics[width=1.1\textwidth]{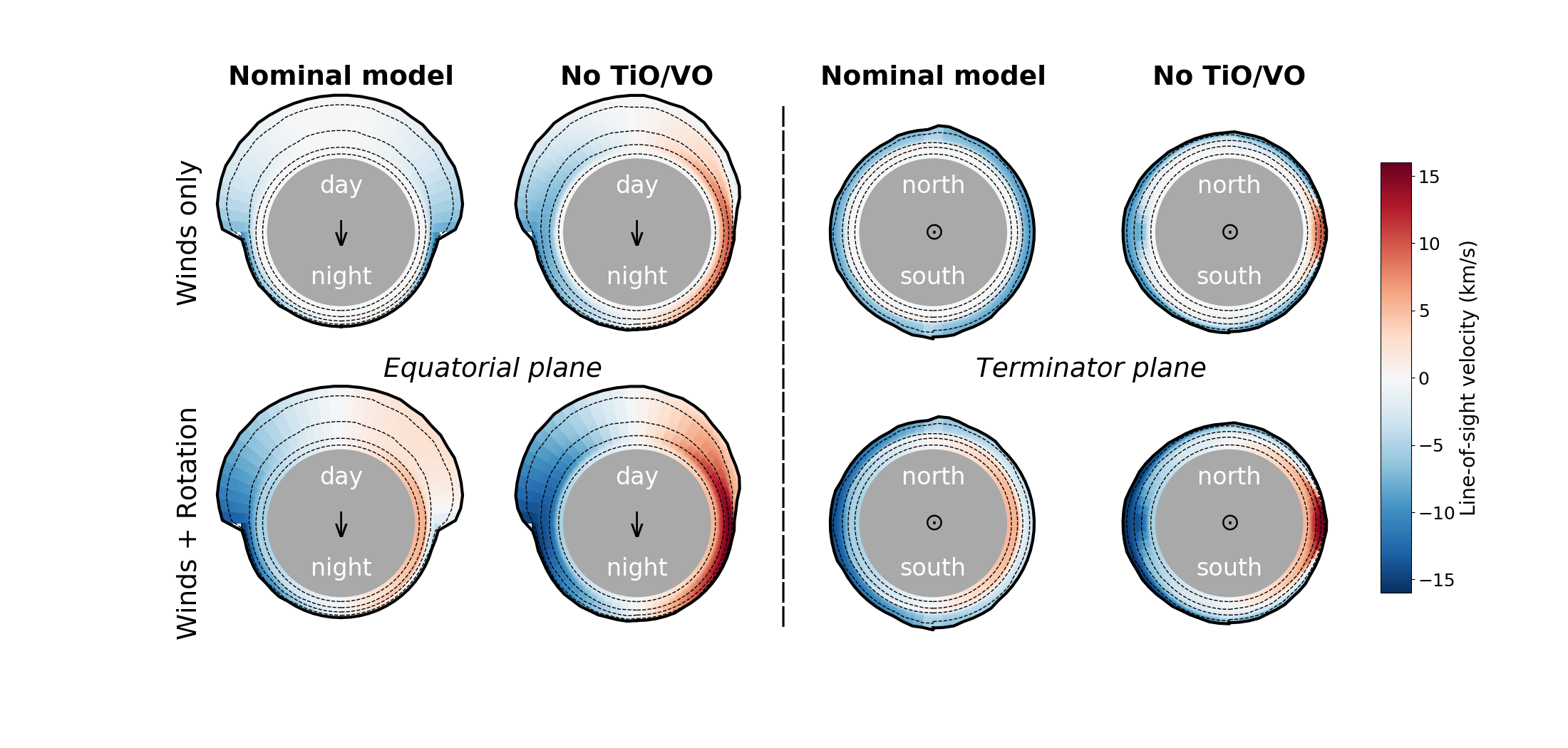}}
\vspace{-45pt}
\caption{Line-of-sight velocities of the nominal model and the no-TiO/VO model at mid-transit, in the equatorial plane (left half of the figure) and the terminator plane (right half). The top row shows the line-of-sight velocities due to \emph{winds only}, while the bottom row shows the line-of-sight velocities that result from the combination of winds and planetary rotation. At the equator, the rotational velocity of WASP-76b is $\sim$5.3 km/s. Black dashes indicate isobars with pressures $10^1$, $10^{-1}$, $10^{-3}$ and $10^{-5}$ bar, respectively (the planet's transmission spectrum probes pressures below $10^{-1}$ bar). {\color{black}{The black arrows indicate the direction of the observer.}}}
\label{fig:winds}
\end{figure*}

\section{Results \& discussion}
\label{sec:results}

\subsection{3D temperatures, abundances, and line-of-sight velocities}
\label{subsec:gcm_results}

Fig. \ref{fig:eq_temps} shows the temperature structure of the four WASP-76b models from Table \ref{tab:gcm_models} in the equatorial plane. As discussed in Section \ref{subsec:model_atmospheres}, the daysides are more ``puffy'' than the nightsides, owing to their higher temperature and lower mean molecular weight. The daysides also feature a strong thermal inversion. For example, at the substellar point in the nominal model, the temperature increases from \mbox{$\sim$1700 K} at \mbox{1 bar} to $\sim$3500 K at 1 mbar. The nightside does not feature a thermal inversion and this is the reason why the cloud deck mostly spans 10 scale heights in the optically-thick-clouds scenario. At the antistellar point, the temperature drops from $\sim$1700 K at 1 bar to $\sim$1000 K at 10 $\mu$bar. 

Fig. \ref{fig:eq_abunds} shows the abundances of Fe, CO, H$_2$O, OH, and TiO across the equatorial plane. All species have a unique 3D spatial distribution. Iron is abundant on the dayside, but absent on the nightside due to condensation. Water, on the other hand, is abundant on the nightside, but subject to thermal dissociation on the dayside (\citealt{Parmentier2018}). These “mirrored” chemical distributions imply that iron lines mainly probe the dayside of the planet, while the water lines mainly probe the nightside. 


The CO abundance is nearly constant across the atmosphere -- its value does not vary by more than $\sim$0.3 dex. Because CO has a strong triple bond between its constituent atoms, it is neither affected by condensation, nor by thermal dissociation. In fact, the only ultra-hot Jupiter hot enough to dissociate CO is KELT-9b (\citealt{Kitzmann2018}). Consequently, the absorption lines of CO are the most reliable gauge of the 3D temperature structure and wind profile of the planet. They only probe spatial variations in temperature and dynamics, and not so much in chemistry. See \citet{Savel2023} for further discussion. 

The distribution of OH is a bit more complicated. On the dayside, the molecule forms when water is dissociated into OH and atomic hydrogen. However, higher up in the atmosphere, OH itself also falls prey to thermal dissociation, producing atomic oxygen and atomic hydrogen. As a result, the OH abundance first increases with altitude and then decreases. On the nightside, OH is absent, because hydrogen and oxygen are contained in water at lower temperatures.   

Finally, TiO is subject to both dissociation on the dayside and condensation on the nightside. Therefore, the only observable TiO is present in a narrow region around the limb where the temperature is \emph{lower} than the dissociation temperature, but \emph{higher} than the condensation temperature of TiO. 



Fig. \ref{fig:winds} shows the line-of-sight velocities $v_\textsc{los}$ due to winds (and planetary rotation) in the equatorial plane and the terminator plane of the nominal and the no-TiO/VO model, at mid-transit. These are the velocities by which the opacities in different cells are Doppler-shifted during the radiative transfer. $v_\textsc{los}<0$ implies that absorbers are moving towards the observer, causing a blueshift to the transmission spectrum, while $v_\textsc{los}>0$ means that absorbers are moving away, inducing a redshift. As illustrated in Fig. \ref{fig:winds}, the nominal model only features day-to-night winds (both planes are completely blueshifted in the top row), such that the only redshift contributions come from rotation (see bottom row). The no-TiO/VO model has an equatorial jet and this is why half of the \emph{equatorial} plane is blueshifted, while the other half is redshifted. Note, though, that the jet only occupies a small region in the \emph{terminator} plane, spanning an angle of $\sim$25 degrees at pressures $\lesssim$ 1 bar on both limbs (the latitudinal extent of the jet is of the order of the equatorial Rossby deformation radius -- see \citealt{Showman2011}). However, despite the smaller ``effective area'' occupied by superrotating winds compared to the day-to-night flow, it may still possible to make inferences about the equatorial jet based on the absorption signal of the full limb (e.g., \citealt{Louden2015}).




\begin{figure*}
\centering
\vspace{0pt} 
\makebox[\textwidth][c]{\hspace{-10pt}\includegraphics[width=0.9\textwidth]{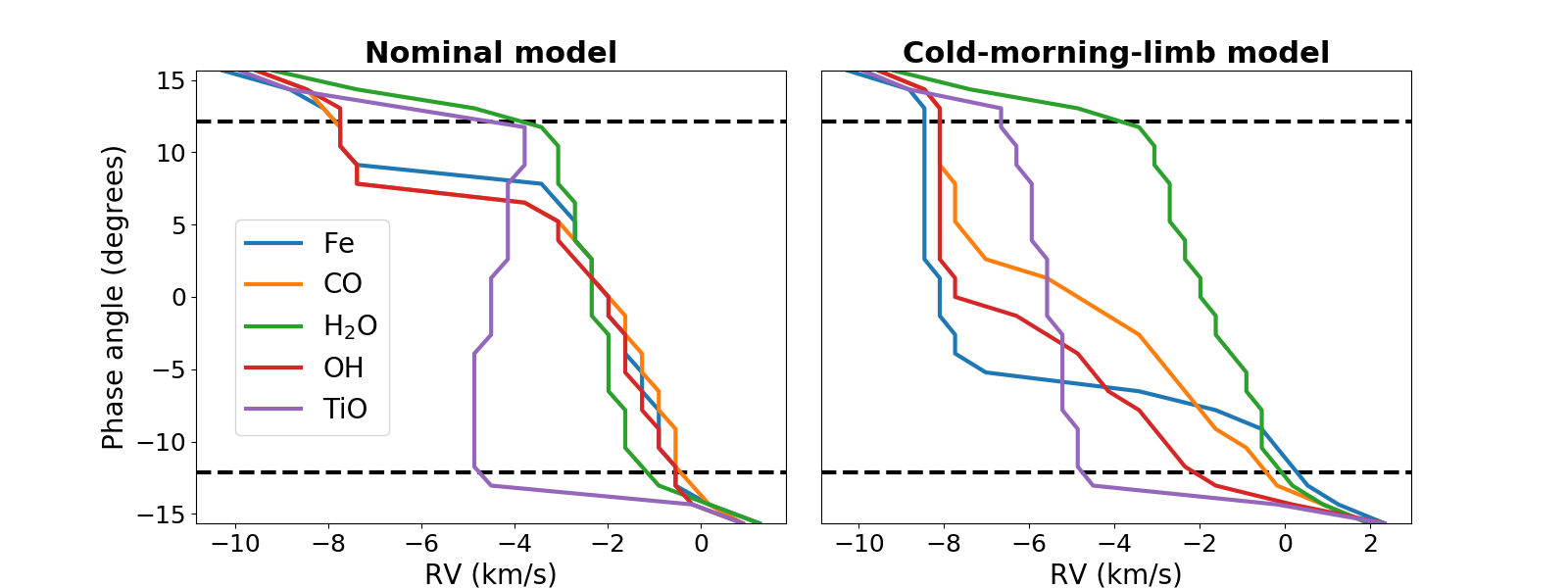}}
\vspace{-10pt}
\caption{The absorption signals of the nominal model (left panel) and the cold-morning-limb model (right panel). For a given orbital phase, the curves show at what radial velocity (RV) the CCF acquires its maximum value. The dashed lines indicate the end of the ingress and the start of the egress, respectively.}
\label{fig:plot_vivien}
\end{figure*}



\subsection{Prelude: the nominal vs. the cold-morning-limb model}

{\color{black}{Fig. \ref{fig:plot_vivien} shows the CCF signals of the nominal model for each of the five chemical species. Remarkably, they all have very similar absorption signatures, except for TiO. However, in the nominal model, the iron signal does not feature the ``kink'' that has been observed in the real data of WASP-76b (\citealt{Ehrenreich2020, Kesseli2021, Pelletier2023} -- see also \citealt{Wardenier2021}). The cold-morning-limb model, on the other hand, does give rise to a kink in the iron signal (blue curve in the right panel of Fig. \ref{fig:plot_vivien}), whereby the blueshift (RV $<$ 0) increases during the first half of the transit and remains constant during the second half. For a full discussion of this behaviour, we refer to \citet{Wardenier2021}.

As opposed to the nominal model, the cold-morning-limb model shows a range of different CCF signals for the five species. In the following sections, our aim is to understand how these CCF signals come about and what physics causes the differences between the models. To build some basic intuition, we start by discussing the CCF maps of the nominal model. Subsequently, we focus on the behaviour of the other three models: the cold-morning limb model, the optically-thick clouds model, and the no-TiO/VO model.}}  

\subsection{CCF maps for the nominal model}
\label{subsec:ccf_maps}

\begin{figure*}
\centering
\vspace{-20pt} 
\makebox[\textwidth][c]{\hspace{10pt}\includegraphics[width=1.1\textwidth]{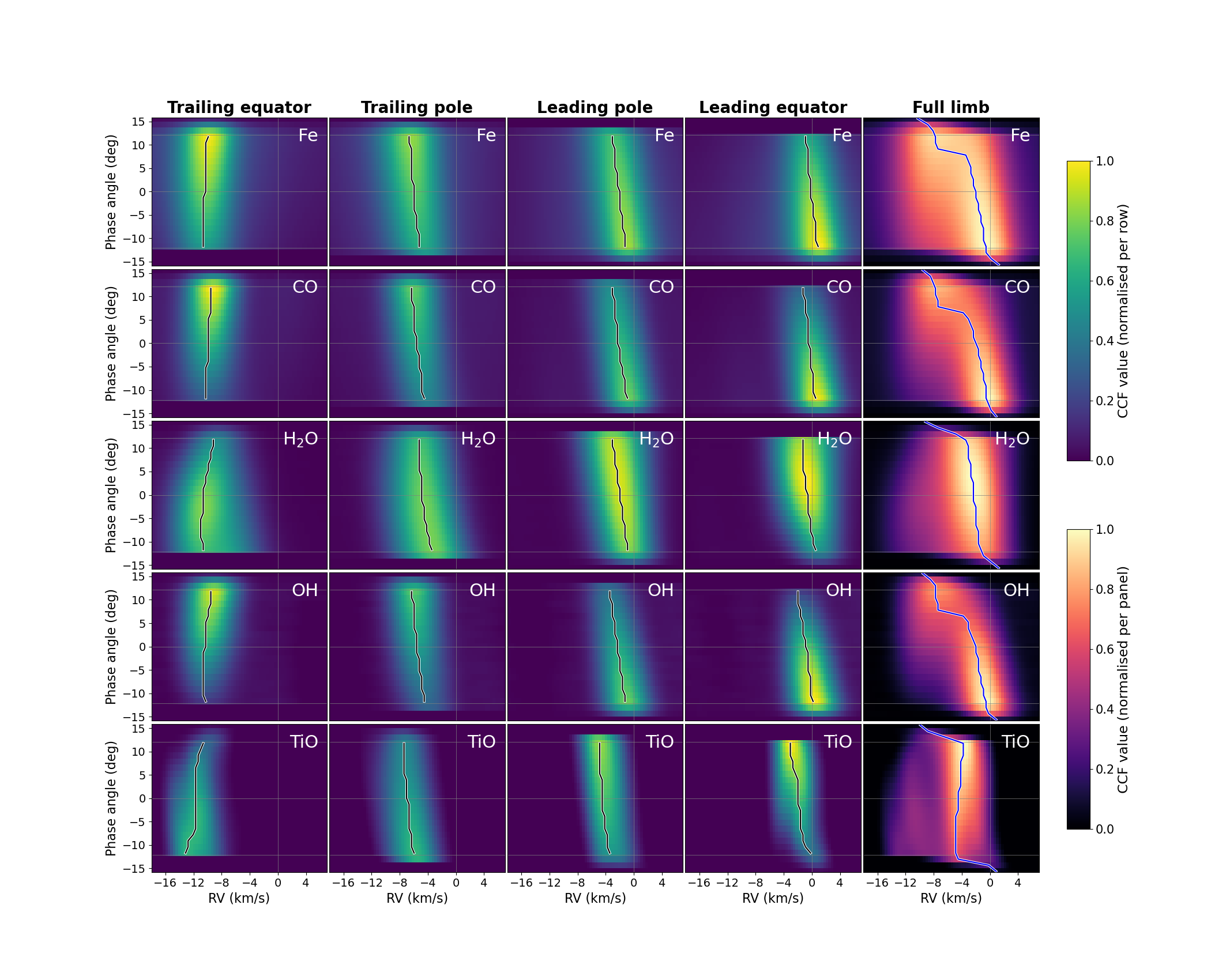}}
\vspace{-45pt}
\caption{CCF maps for each of the limb sectors of the nominal model (first four columns), for each of the species considered in this work (rows). The right column contains the CCF maps of the full limb. For each species, the CCF maps of the limb sectors are normalised to the overall maximum of the four maps. The CCF maps of the full limb are normalised to their own maximum. The solid line in each panel marks the maximum value of the CCF as a function or orbital phase.}
\label{fig:all_sectors}
\end{figure*}

\begin{figure*}
    \centering

    \begin{minipage}[c]{\textwidth}
    \makebox[\textwidth][c]{
    \includegraphics[width=1.2\textwidth]{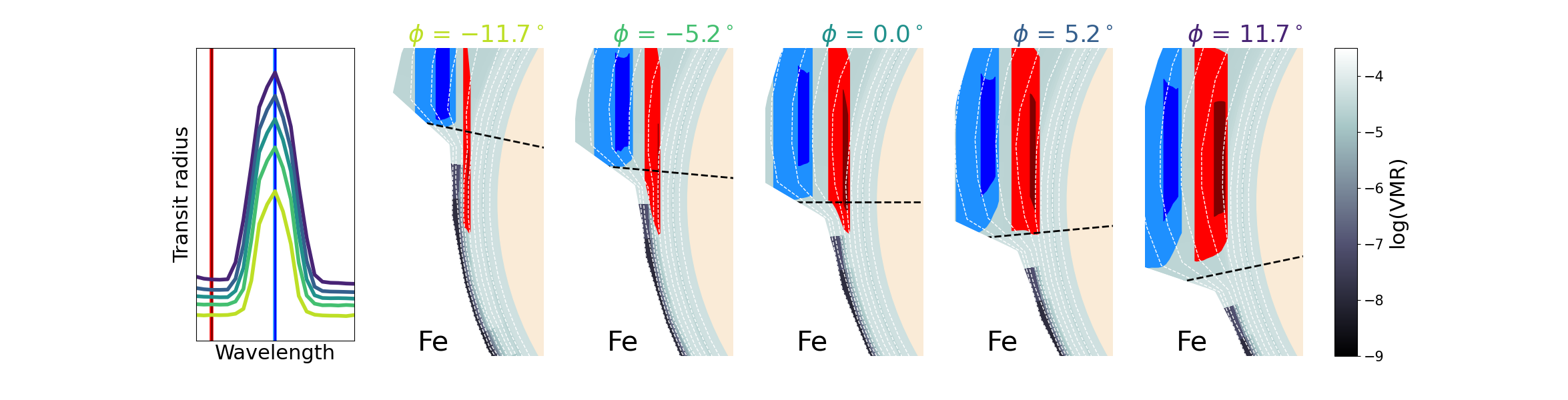}}
    \end{minipage} 

    \vspace{-10pt}
    
    \begin{minipage}[c]{\textwidth}
    \makebox[\textwidth][c]{
    \includegraphics[width=1.2\textwidth]{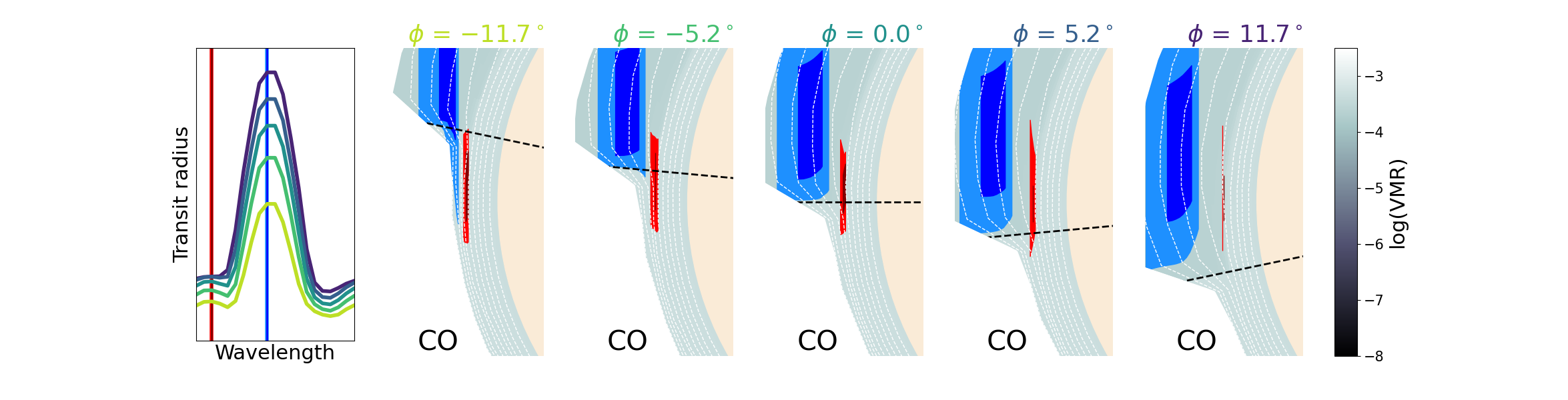}}
    \end{minipage}

    \vspace{-10pt}

    \begin{minipage}[c]{\textwidth}
    \makebox[\textwidth][c]{
    \includegraphics[width=1.2\textwidth]{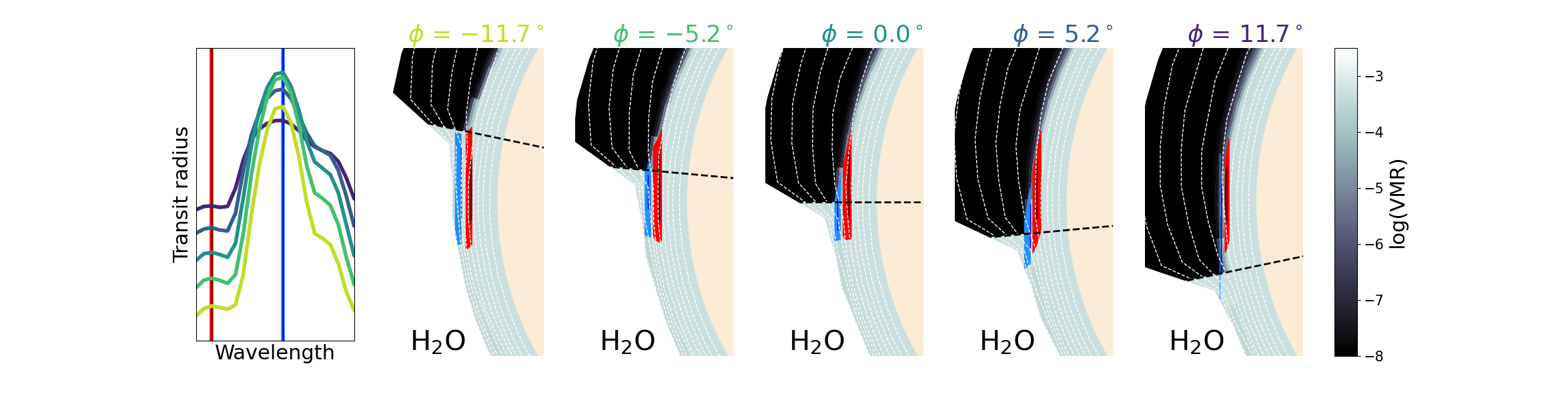}}
    \end{minipage} 

     \vspace{-7pt}

    \begin{minipage}[c]{\textwidth}
    \makebox[\textwidth][c]{
    \includegraphics[width=1.35\textwidth]{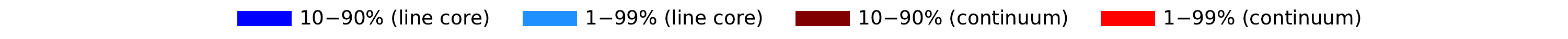}}
    \end{minipage} 

    \vspace{-25pt}

    \begin{minipage}[c]{1\textwidth}
      \vspace{25pt}
	  \caption{Absorption regions in the trailing part of the equatorial plane of the nominal model, for Fe (top row), CO (middle row), and H$_2$O (bottom row). For all species, absorption regions are plotted at two wavelengths, one inside a line core (blue) and one in the continuum (red). The 10--90$\%$ and 1--99$\%$ absorption regions are visualised at five different orbital phase angles ranging from $-11.7^\circ$ (just after ingress) to $+11.7^\circ$ degrees (just before egress). In each panel, the colourmap depicts the abundance of the relevant species, while the black dashed line marks the terminator plane. The dayside lies ``above'' this line and the nightside lies ``below'' this line. Furthermore, the white dashes indicate isobars that go down from 10 to $10^{-7}$ bar, in steps of 1 dex. The left panels show the absorption lines of the trailing equator at the five orbital phases (note that these only contribute for 25$\%$ towards the full spectrum).} \label{fig:absorption_regions_all_1}
    \end{minipage}
\end{figure*}

Fig. \ref{fig:all_sectors} depicts the CCF maps of the four limb sectors and the full limb of the \emph{nominal model}, for all species. The CCF maps of the individual sectors can be seen as the ``building blocks'' of the more complicated absorption signal that emerges from the entire atmosphere. This is because the the CCF map of the full limb is the sum of the maps of the limb sectors. Figs. \ref{fig:absorption_regions_all_1} and \ref{fig:absorption_regions_all_2} show the absorption regions that are probed on the trailing part of the equatorial plane by (randomly chosen) line cores of Fe, CO, H$_2$O, OH, and TiO, respectively.

\subsubsection{Recap: two important effects for iron}

\citet{Wardenier2021} {\color{black}{(see their Fig. 9)}} showed that there are two important effects that drive the Doppler shift of iron in the nominal model: (i) the variation in the signal strengths of the limb sectors during the transit, and (ii) atmospheric dynamics. In tandem, these effects cause the absorption signal to become \emph{increasingly} blueshifted during the transit, even though there is \emph{no} significant thermal or chemical asymmetry between the planet's trailing and leading limbs. On top of this ``baseline behaviour'', limb asymmetries (e.g., \citealt{Savel2023}) can further enhance changes in the Doppler shift with orbital phase (see Section \ref{subsec:model_comparisons}). 

Effect (i) is due to the day-night temperature contrast of the planet, in combination with tidally-locked rotation. Ignoring the contribution from winds, the absorption signal of the redshifted leading limb becomes \emph{weaker} during the transit, as the dayside rotates out of view. On the other hand, the signal of the blueshifted trailing limb becomes \emph{stronger}, as the dayside rotates into view. Therefore, in a scenario without winds, the absorption signal of the full limb transitions from being mainly redshifted in the first half of the transit to being mainly blueshifted in the second half\footnote{See the third row of Fig. 9 in \citet{Wardenier2021} (nominal model w/o winds).}. 

Effect (ii), atmospheric dynamics, impacts the absorption signal of the planet in multiple ways. Firstly, winds shift the whole CCF to negative radial velocities, as the day-to-night flow causes the whole terminator plane to be blueshifted (see Fig. \ref{fig:winds}). Secondly, they ``smoothen'' the CCF map, resulting in a more gradual change of the net Doppler shift as a function of orbital phase. Finally, the angle between the polar wind vectors and the line-of-sight vector becomes smaller during the transit (see Fig. 10 in \citealt{Wardenier2021}). This projection effect causes the absorption signals associated with the polar sectors to become more blueshifted over time. The signal of the leading equator shows the same behaviour. 

\subsubsection{Fe signals}



The first row of Fig. \ref{fig:all_sectors} shows the CCF maps for iron. As mentioned in the previous paragraph, the iron signals of the nominal model were already presented in \citet{Wardenier2021}, but not for the entire ESPRESSO wavelength range. Although the atmosphere does not feature strong limb asymmetries, the iron lines become progressively more blueshifted during the transit, owing to the effects discussed in the previous section. 



\citet{Wardenier2021} {\color{black}{(see their Section 5.1)}} suggested that the change in signal strength of the limb sectors was because the observation first probes the nightside on the trailing limb, and later the dayside -- with the exact opposite occurring on the leading limb. However, Fig. \ref{fig:absorption_regions_all_1} demonstrates that something else is going on. Inside an iron line core, the absorption region lies on the dayside \emph{at every orbital phase}. The reason why the signal of the trailing limb becomes stronger, though, is the fact that the projected separation (onto the limb plane) between the absorption region of the line core (in blue) and the absorption region of the continuum (in red) becomes larger during the transit. As shown in the top-left panel of Fig. \ref{fig:absorption_regions_all_1}, this causes an iron line to become stronger relative to its continuum, which is exactly what the magnitude of the CCF encodes. Therefore, in the case of iron, changes to a sector's signal strength \emph{are a consequence of geometry}, rather than absorption regions shifting between the dayside and the nightside of the planet. However, the day-night contrast is still crucial for the projection effect to occur.

Furthermore, Fig. \ref{fig:absorption_regions_all_1} illustrates that if iron was uniformly distributed across the atmosphere, its absorption lines would still only probe the dayside (see also the next paragraph about CO). Hence, it is \emph{not} the 3D chemical map of iron that confines its absorption regions to the dayside. Instead, the scale-height difference between the dayside and the nightside causes a ``shielding effect'' -- because the dayside is more puffy, the absorption lines probe altitudes at which the opacity of the nightside is negligible.

\subsubsection{CO signals}

The second row of Fig. \ref{fig:all_sectors} shows the CCF maps of CO. Both the CCF maps and the absorption regions of CO (see Fig. \ref{fig:absorption_regions_all_1}) display nearly identical behaviour compared to iron. Although the abundance of CO is uniform across the atmosphere, its absorption lines virtually only probe the dayside. That is, the 10$-$90\% absorption regions of the CO line core are situated on the dayside at all orbital phases. The reason for this is the ``shielding effect'' discussed in the previous paragraph. Stellar light rays first encounter the dayside, and the $\tau$$\sim$1 region lies at altitudes where the nightside does not contribute to the total optical depth.

For CO, the signal strength of the trailing equator also increases during the transit, again due to a projection effect. Around the CO line plotted in Fig. \ref{fig:absorption_regions_all_1}, the ``continuum'' is caused by water absorption. As a result, the absorption region of the continuum behaves exactly like that of water in Fig. \ref{fig:absorption_regions_all_1} (see also the next paragraph). Interestingly, at $\phi = -11.7^\circ$ the CO line core and its adjacent continuum probe \emph{different} sides of the planet.

\begin{figure*}
\centering
\vspace{-35pt} 
\makebox[\textwidth][c]{\hspace{10pt}\includegraphics[width=1.1\textwidth]{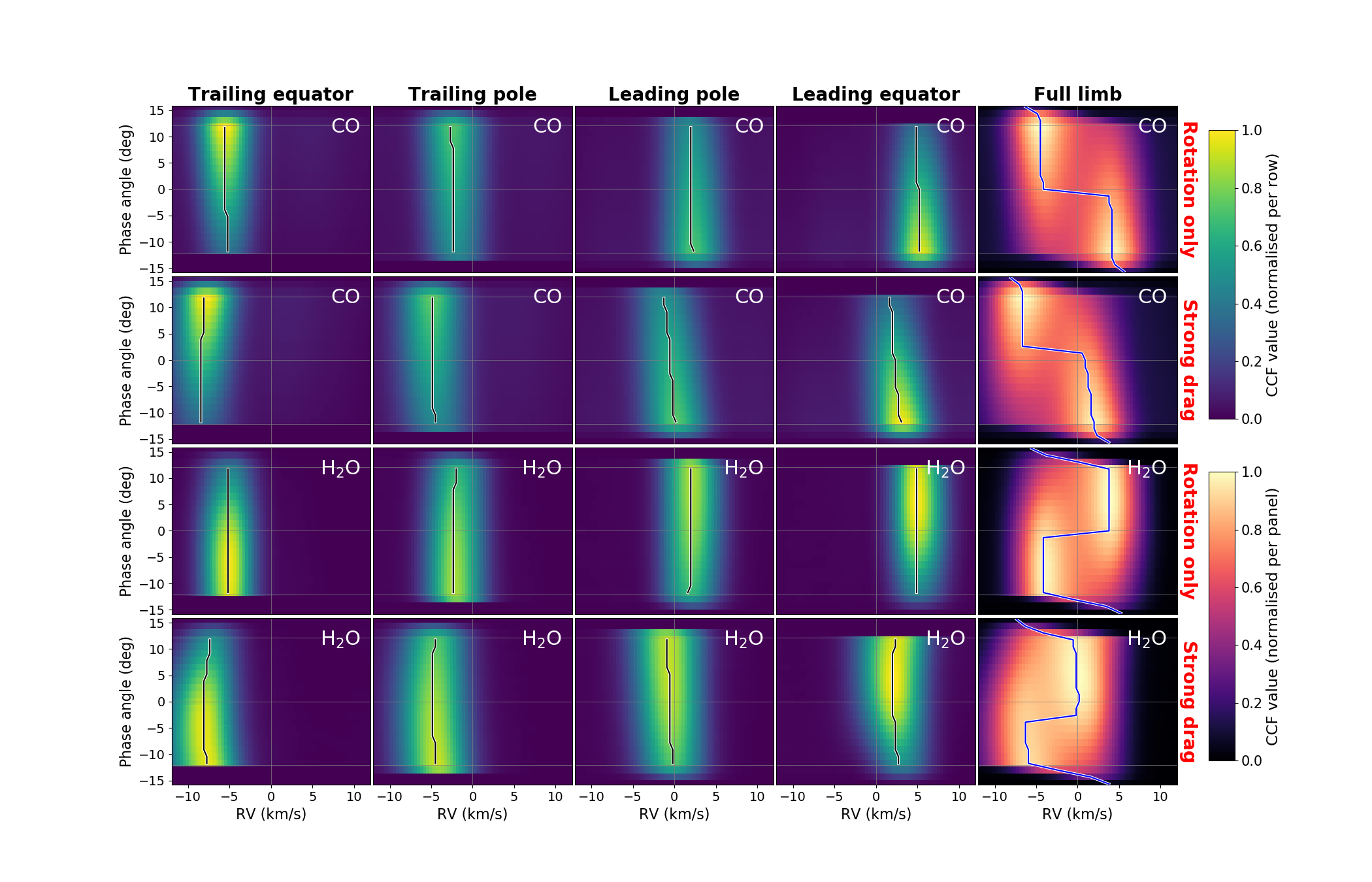}}
\vspace{-35pt}
\caption{{\color{black}{CCF maps for CO (top rows) and H$_\text{2}$O (bottom rows) for two additional realisations of the nominal model. In the ``rotation only'' scenario, we only account for Doppler shifts due to planet rotation, switching off the winds in gCMCRT. In the ``strong drag'' scenario, the drag timescale $\tau_\text{drag}$ in the GCM is reduced from $10^5$ to $10^4$ s. As a consequence, winds are slowed down and planet rotation becomes the more dominant contributor to the line-of-sight velocities. Note that the horizontal axis of the panels has a different scale compared to Fig. \ref{fig:all_sectors}.}}}  
\label{fig:extra_sector_plot}
\end{figure*}


\subsubsection{H$_\text{2}$O signals}

The third row of Fig. \ref{fig:all_sectors} shows the CCF maps of water. Remarkably, the phase-dependence of the signal strengths displays the opposite behaviour compared to iron and CO. For iron and CO, the trailing-equator signal becomes \emph{stronger} during the transit, but for water it becomes \emph{weaker}. The absorption regions in Fig. \ref{fig:absorption_regions_all_1} demonstrate why this is the case. At the start of the transit, a water line core probes the nightside. Then, as the planet rotates, its absorption region shifts towards the dayside. However, because of the lack of water at low pressures (due to thermal dissociation), the absorption region is ``pushed'' down to higher pressures on the dayside. At these higher pressures, the absorption regions of the line core and the continuum lie closer together, which explains why the signal strength of the trailing limb decreases over the transit. Naturally, the opposite occurs on the leading limb, as shown by the CCF maps of the limb sectors in Fig. \ref{fig:all_sectors}.

\begin{figure*}
    \centering
    
    \begin{minipage}[c]{\textwidth}
    \makebox[\textwidth][c]{
    \includegraphics[width=1.2\textwidth]{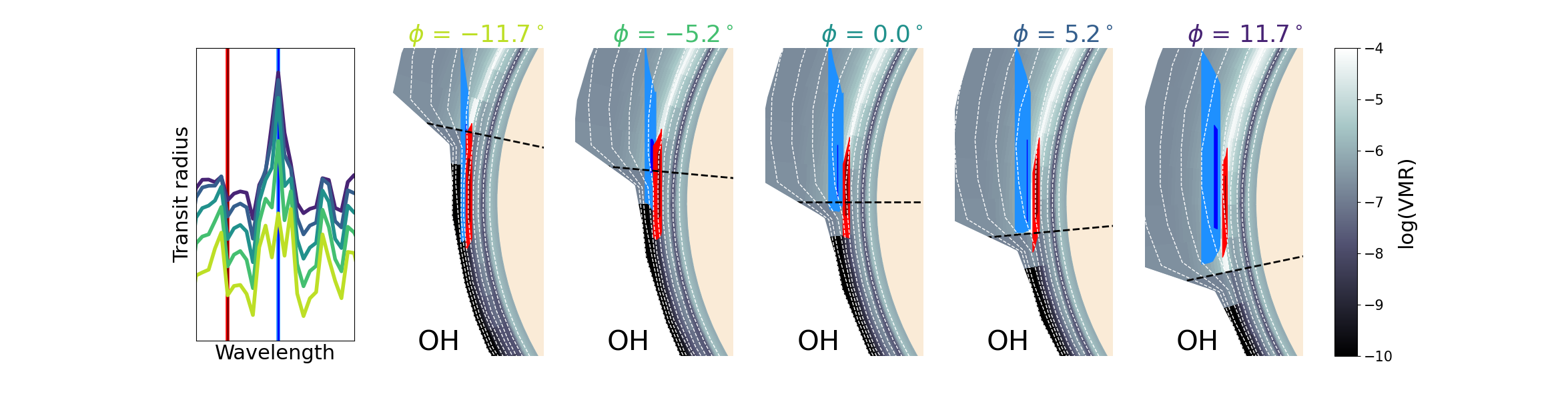}}
    \end{minipage}

    \vspace{-10pt}

    \begin{minipage}[c]{\textwidth}
    \makebox[\textwidth][c]{
    \includegraphics[width=1.2\textwidth]{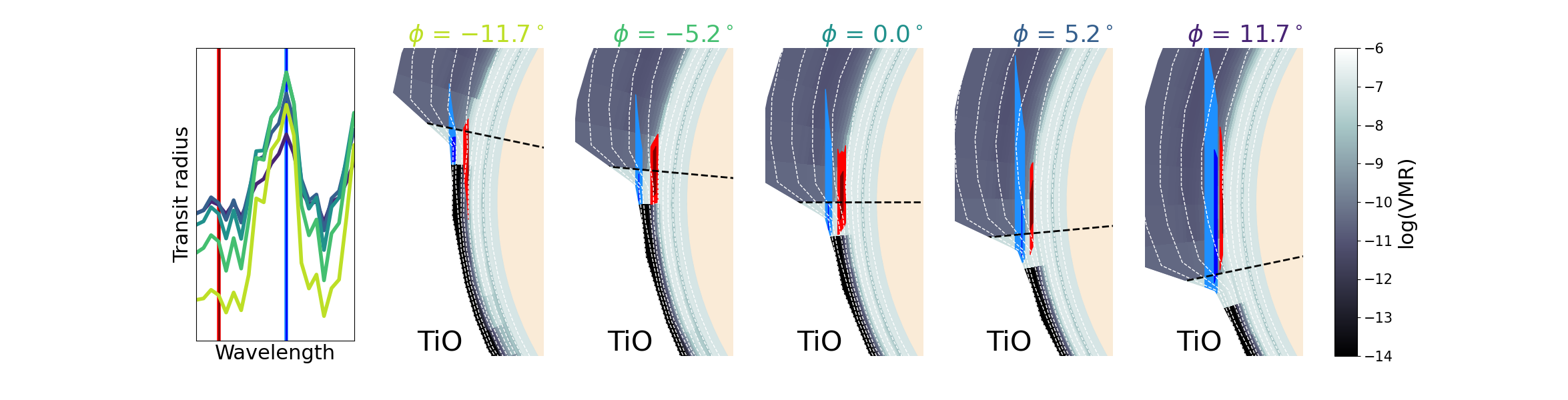}}
    \end{minipage} 

     \vspace{-7pt}

    \begin{minipage}[c]{\textwidth}
    \makebox[\textwidth][c]{
    \includegraphics[width=1.35\textwidth]{figures/A_legend.pdf}}
    \end{minipage} 

    \vspace{-25pt}

    \begin{minipage}[c]{1\textwidth}
      \vspace{25pt}
	  \caption{Same as Fig. \ref{fig:absorption_regions_all_1}, but for OH (top row) and TiO (bottom row).} \label{fig:absorption_regions_all_2}
    \end{minipage}
\end{figure*}

Based on the behaviour of the individual limb sectors, one would expect the blueshift of the full limb to \emph{decrease} during the transit. After all, the signal strength of the (least blueshifted) leading sectors becomes stronger over time. However, the reason why the net Doppler shift of the planet does \emph{not} decrease is that the signals of the leading limb are stronger than those of the trailing limb over almost the entire transit. Therefore, the water signal of the full limb is dominated by the leading sectors, which are subject to an \emph{increasing} blueshift over time. 

{\color{black}{To build more physical understanding, Fig. \ref{fig:extra_sector_plot} shows the CCF maps of CO and water for two additional realisations of the nominal model: (i) a scenario with rotation only, in which the winds are zero, and (ii) a scenario with a shorter drag timescale \mbox{$\tau_\text{drag}$ = $10^4$ s}, in which the winds are weaker (as they are subject to stronger drag) compared to the original model with \mbox{$\tau_\text{drag}$ = $10^5$ s}. In the rotation-only case, the Doppler shifts of the individual limb sectors are \emph{constant} during the transit. This means that the phase-dependence of the absorption trail of the full limb is purely governed by the \emph{varying signal strengths} of the limb sectors. Therefore, in the rotation-only case, we do see that CO and water display the exact opposite behaviour: the CO signal goes from redshifted \mbox{(RV $>$ 0 km/s)} to blueshifted \mbox{(RV $<$ 0 km/s)}, while the water signal goes from blueshifted to redshifted.}} {\color{black}{In the scenario with strong drag (second and fourth row in Fig. \ref{fig:extra_sector_plot}), planet rotation still dominates the shape of the absorption signals of the full limb. However, the signals are now more blueshifted because of the prevalence of day-to-night winds. 

As shown in Fig. \ref{fig:all_sectors}, increasing the drag timescale (\mbox{$\tau_\text{drag}$ = $10^4$ s $\rightarrow$} \mbox{$\tau_\text{drag}$ = $10^5$ s}) causes the ``step'' feature in the planet's water signal to disappear. As previously mentioned, this is because the leading-limb signal becomes stronger than the trailing-limb signal over the entire transit. Hence, it is the planet's 3D wind-profile that is inducing an asymmetry in the nominal model: on the trailing limb, the variance in probed wind speeds is larger, causing the line contrast to become smaller and the CCF of the trailing sectors to become broader. Ultimately, this causes the water signal to look relatively similar to that of iron and CO, even though the water lines probe completely different regions of the atmosphere. Our result is in qualitative agreement with \citet{Savel2023} and \citet{Beltz2023}, who also found an increasing blueshift for water with their ``baseline'' 3D models of WASP-76b. Furthermore, \citet{Beltz2023} also reported a step feature in the water signal of one of their magnetic-drag models, hinting at weaker winds and a more visible signature of planet rotation.}} 


On a final note, the absorption regions of water become very narrow towards the end of the transit ($\phi = 11.7^\circ$ in Fig. \ref{fig:absorption_regions_all_1}). This is because the distance between isobars is smaller at higher pressures. Also, the absorption regions coincide with a steep vertical gradient in the water abundance. Therefore, shifting the transit chord to a lower pressure will result in a sharp decrease in integrated abundance (and thus optical depth $\tau$), while moving it to higher pressures will result in $\tau \gg 1$.

\begin{figure*}
\centering
\vspace{-40pt} 
\makebox[\textwidth][c]{\hspace{10pt}\includegraphics[width=1.1\textwidth]{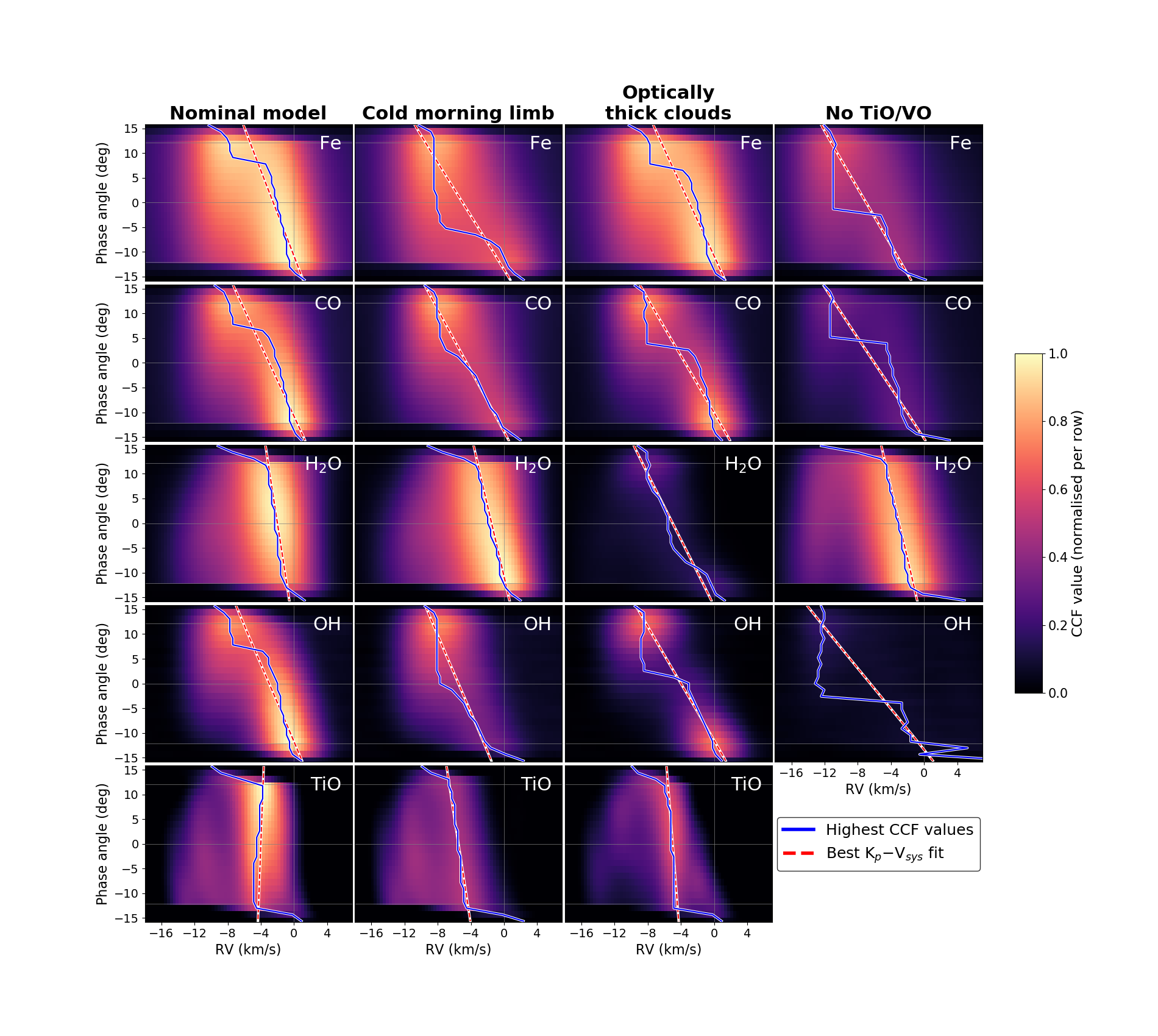}}
\vspace{-45pt}
\caption{CCF maps of Fe, H$_2$O, CO, OH, and TiO (one species species per row) for each of the four models considered in this work (see Table \ref{tab:gcm_models} and Figs. 2--4). All maps were normalised to their own maximum. The colourmap shows the value of the cross-correlation as a function of orbital phase (vertical axis) and velocity shift between the template and the transmission spectrum of the planet (horizontal axis). The blue curve in each panel indicates the maximum value of the CCF at each orbital phase. The red dashed line corresponds to the $K_\text{p}$ and $V_\text{sys}$ values that give rise to the maximum SNR values in the $K_\text{p}$--$V_\text{sys}$ map.}
\label{fig:all_ccfs}
\end{figure*}


\subsubsection{OH signals}

The fourth row of Fig. \ref{fig:all_sectors} shows the CCF maps of OH. In the nominal model, the OH signal resembles that of iron and CO. However, the strength of the OH signal drops by a factor $\sim$2 during the transit. Also, Fig. \ref{fig:all_sectors} demonstrates that the change in the signal strength of the trailing equator (as well as the other limb sectors) is more extreme than for the other species -- at the start of the transit it is almost zero. Fig. \ref{fig:absorption_regions_all_2} illustrates the cause of this significant variation. Because the nightside is depleted of OH and because the higher-altitude regions on the dayside have a low OH abundance, the absorption regions of the line core and the continuum overlap at the start of the transit. This produces a negligible CCF signal. However, as the planet rotates, the more OH-abundant parts of the dayside rotate into view, and the line contrast increases.  

\subsubsection{TiO signals}

The bottom row of Fig. \ref{fig:all_sectors} shows the CCF maps of TiO. Whereas the other species clearly show an increasing blueshift during the transit, the blueshift of TiO is \emph{decreasing} in the nominal model (albeit marginally). The reason for this is that the signal strengths of the limb sectors behave like those of water (see Fig. \ref{fig:all_sectors}). Yet, in contrast to water, the signal of the trailing sectors is strong enough at the beginning to contribute to the Doppler shift of the full limb. At the end of the transit, the CCF map is dominated by the signal from the leading sectors.

As shown in the figures, the signal of the (most blueshifted) trailing limb becomes \emph{weaker} during the transit, as the absorption region of the TiO line core shifts from the nightside to the dayside. We note, though, that the absorption region very much hinges on the TiO abundances in the first column on the nightside (see Fig. \ref{fig:absorption_regions_all_2}), where the temperature profile allows for TiO to exist at all pressures. Without this column, the absorption regions would have been situated at lower altitudes, resulting in much weaker absorption lines (the column \emph{has} to exist, though, as the atmosphere transitions from dayside to nightside). On the other hand, if the temperature gradient at the terminator was smoother, for example due to H$_2$ dissociation/recombination (\citealt{Bell2018,Komacek2018,Tan2019,Roth2021}), TiO would have existed across a wider range of longitudes. Therefore, in this model, it is the steepness of the temperature gradient at the terminator that determines whether or not TiO may be observable.


\subsection{CCF maps for other models}
\label{subsec:model_comparisons}

Fig. \ref{fig:all_ccfs} shows the CCF maps computed for all models and all species (only the CCF maps of the full limb are shown). The colourmaps were normalised per row to allow for inter-model comparisons. Because TiO is cold-trapped in the no-TiO/VO model, it is not observable. Hence, there are 19 maps in total, rather than 20. Fig. \ref{fig:plot_vivien} depicts the absorption trails of all species in the same panel, for the nominal model and the cold-morning-limb model.

\subsubsection{Fe signals}

As expected, the cold-morning-limb model in Fig. \ref{fig:all_ccfs} shows a strong increase in blueshift over the first half of the transit. The reason for this is that the signals of the leading sectors are much weaker compared to the nominal model (see Fig. 13 in \citealt{Wardenier2021}). Hence, the signals of the trailing sectors already start to dominate the sum before mid-transit. After mid-transit, the blueshift remains constant, because the only contributions to the signal come from the trailing sectors. For further discussion, we refer to \citet{Wardenier2021}.

The CCF map of the optically-thick-clouds model is very similar to that of the nominal model, suggesting that adding a cloud deck has minimal impact on the iron signal. The reason for this is that the absorption regions of iron on the dayside lie at much \emph{higher} altitudes compared to the optically thick clouds on the nightside (see Figs. \ref{fig:eq_abunds} and \ref{fig:absorption_regions_all_1}). Yet, the signal strength of the full limb does decrease slightly as a result of clouds. This is because the cloud deck lies at higher altitudes than the absorption regions of the continuum in the cloud-free atmosphere. Since the cloud deck is nearly symmetric, the signals of the trailing and leading limbs are affected equally -- the line contrast and the magnitude of the CCF become marginally smaller, but the shape does not change. Thus, in this particular model, we find that nightside clouds are unable to mute the iron absorption signal, as it originates from too high altitudes. To make a cloud mute the absorption features of iron, as in \citet{Savel2022}, it should be located at a significantly higher altitude than the continuum in the cloud-free case. {\color{black}{Hence, based on our modelling efforts, we still find that a temperature (or scale-height) asymmetry between the trailing and leading limb is the most likely explanation for the strongly blueshifted iron signals of WASP-76b}} (\citealt{Ehrenreich2020,Kesseli2021,Pelletier2023}) and WASP-121b (\citealt{Borsa2021}). 


The iron signal of the no-TiO/VO model also exhibits stronger blueshifts compared to the nominal model. This is also related to a temperature asymmetry (see Fig. \ref{fig:eq_temps}). Due to a large hotspot shift on the dayside, the 3D temperature structure of the model is lopsided, with the trailing limb being hotter and more extended than the leading limb. Hence, the signal of the blueshifted trailing limb contributes more strongly to the CCF map of the full planet.

\subsubsection{CO signals}

The behaviour of CO across the different models is very similar to that of iron (e.g., compare the first and the second rows in Figs. \ref{fig:all_sectors} and \ref{fig:all_ccfs}). In the cold-morning-limb model, for example, the signal also undergoes a strong increase in blueshift during the first half of the transit, owing to the temperature asymmetry between the trailing and the leading sectors. 

The CCF map of CO is somewhat affected by the presence of optically thick clouds. This indicates that there are weaker CO lines that probe the atmosphere at \emph{lower} altitudes, and which are thus muted by the cloud deck. For these weaker lines, the absorption regions are likely to lie partly on the dayside and partly the nightside, as CO is equally abundant on both hemispheres. With this in mind, the (stronger) CO line considered in Fig. \ref{fig:absorption_regions_all_2} may not be fully representative. However, note that the blue absorption regions plotted in Fig. \ref{fig:absorption_regions_all_2} only pertain to the line \emph{core} -- the line wings, which also contribute to the CCF, must probe lower altitudes.

\subsubsection{H$_2$O signals}

The signals of the nominal model, the cold-morning-limb model, and the no-TiO/VO model show the same behaviour, with the no-TiO/VO signal being slightly more blueshifted due to stronger day-to-night winds. The reason for this is that the 3D spatial distribution of water in each of the models is very similar. Water is present on the nightside, as well as at higher pressures on the dayside where the scale heights on the trailing limb and the leading limb are still the same \mbox{(see Fig. \ref{fig:eq_abunds})}. Consequently, the temperature asymmetries in the cold-morning-limb model and the no-TiO/VO model do not manifest in the CCF maps.

In contrast to iron, the presence of optically thick clouds strongly suppresses the water signal. This is because the cloud deck is situated at roughly the same altitude as the water absorption regions. Hence, the line contrast is small and the vast majority of water lines are muted.

\begin{figure*}
\centering
\vspace{-40pt} 
\makebox[\textwidth][c]{\hspace{10pt}\includegraphics[width=1.1\textwidth]{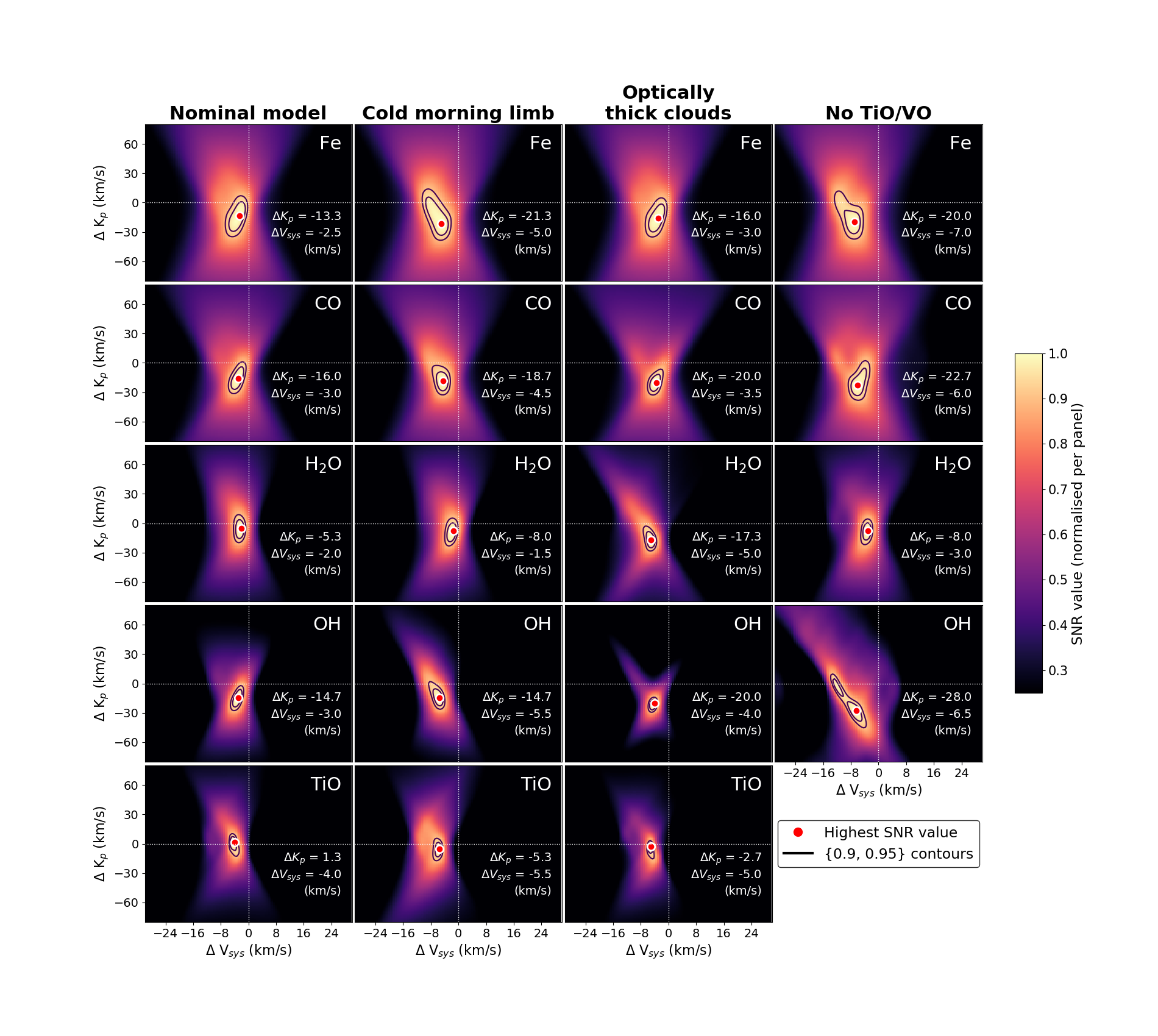}}
\vspace{-45pt}
\caption{$K_\text{p}$--$V_\text{sys}$ maps of Fe, CO, H$_2$O, OH, and TiO (one species species per row) for each of the four models considered in this work (see Table \ref{tab:gcm_models} and Figs. 2--4). All maps were normalised to their own maximum. The colourmap shows the value of the SNR (signal-to-noise ratio), as a function of orbital velocity (vertical axis) and systemic velocity (horizontal axis) \emph{relative to the planetary rest frame}. The red marker in each panel indicates the maximum SNR value \mbox{(SNR = 1)}, while the black contours correspond to SNR = 0.9 and SNR = 0.95, respectively.}
\label{fig:all_kpvsys}
\end{figure*}


\subsubsection{OH signals}

The OH signal of the cold-morning-limb model also features an increasing blueshift during the first half of the transit. However, the signal from the leading sectors is very weak. The reason for this is that the colder leading limb is depleted of OH, while the high-altitude part of the dayside that is in view does hot have sufficient OH abundance to cause significant absorption. In the second half of the transit, the trailing sectors completely dominate the absorption signal. The same idea holds for the no-TiO/VO model, which does not have any detectable OH on the leading limb.

Just like water, OH probes relatively low altitudes. Therefore, the introduction of optically thick clouds heavily mutes the OH absorption lines. At mid-transit, the absorption signal is almost zero. The strongest contributions come from the leading equator around ingress and from the trailing equator around egress. Again, however, it is questionable whether optically thick clouds allow for OH to be detected at all, given that the CCF signal only emerges from a narrow range of orbital phase angles.


\subsubsection{TiO signals}

In the cold-morning-limb model, the contribution from the leading equator is zero (not shown in a plot). This is why the TiO signal is more blueshifted than in the nominal model. Additionally, the TiO signal appears to have a more bimodal structure. As a result, the signal is ``smeared'' over a range of $K_\text{p}$ and $V_\text{sys}$ values (see Fig. \ref{fig:all_kpvsys}), which could make TiO harder to detect in this scenario.

When optically thick clouds are introduced, TiO absorption is muted in all limb sectors except the leading pole (not shown). This is why the blueshift of the full-limb signal in Fig. \ref{fig:all_ccfs} closely resembles that of the leading pole in Fig. \ref{fig:all_sectors}.

\subsection{$K_\text{p}$--$V_\text{sys}$ maps}
\label{subsec:kpvsys_maps}

\subsubsection{Systematic peak offsets}

Fig. \ref{fig:all_kpvsys} shows the $K_\text{p}$--$V_\text{sys}$ maps associated with the CCF maps from Fig. \ref{fig:all_ccfs}. Because the absorption signals all exhibit Doppler shifts in the planetary rest frame, the SNR peaks in the $K_\text{p}$--$V_\text{sys}$ maps are offset from (0, 0) km/s. All SNR peaks, except for the the TiO signal of the nominal model, are consistently located at \emph{lower} $V_\text{sys}$ and \emph{lower} $K_\text{p}$ values than would be expected based on the orbital motion of the planet and the radial velocity of the star. The red dashed curves in Fig. \ref{fig:all_ccfs} illustrate why this is the case. These are the curves that give rise to the highest integrated SNR value (equation \ref{eq:kpvsys_sum}). Because all absorption signals are blueshifted on average, the best-fitting curve has a negative horizontal offset, yielding $\Delta V_\text{sys} < 0$. Also, because the signals become \emph{more} blueshifted over time (with the exception of TiO in the nominal model), the slope of the curve is negative, corresponding to $\Delta K_\text{p} < 0$.

Along the $V_\text{sys}$ axis, the offset of the SNR peak is typically a few km/s. To zeroth order, $\Delta V_\text{sys}$ can be interpreted as the average wind speed across the terminator. The maximum shift we encounter is $\Delta V_\text{sys} = -7$ km/s for the iron signal of the no-TiO/VO model. Intuitively, it makes sense that the shifts are the largest for this model, as it has no drag and thus the highest wind speeds. 

Along the $K_\text{p}$ axis, the peak offset can be more significant. Typically, the $K_\text{p}$ shift is much larger than the Doppler shift measured from the CCF at any orbital phase. This is because the value of $\Delta K_\text{p}$ does \emph{not} encode information about the \emph{absolute value} of the line-of-sight velocities. Rather, it reflects the \emph{rate of change} of the planet's Doppler shift during the transit (and thus how steep the slope of the best fitting absorption trail needs to be). For this reason, signals with a strong phase-dependence show the most extreme values of $\Delta K_\text{p}$. For example, both the cold-morning-limb model and the no-TiO/VO model yield $\Delta K_\text{p} \approx -20$ km/s for iron and CO.

\begin{figure}
\centering
\vspace{-10pt}
\makebox[\textwidth][c]{
\hspace{-250pt}
\includegraphics[width=0.54\textwidth]{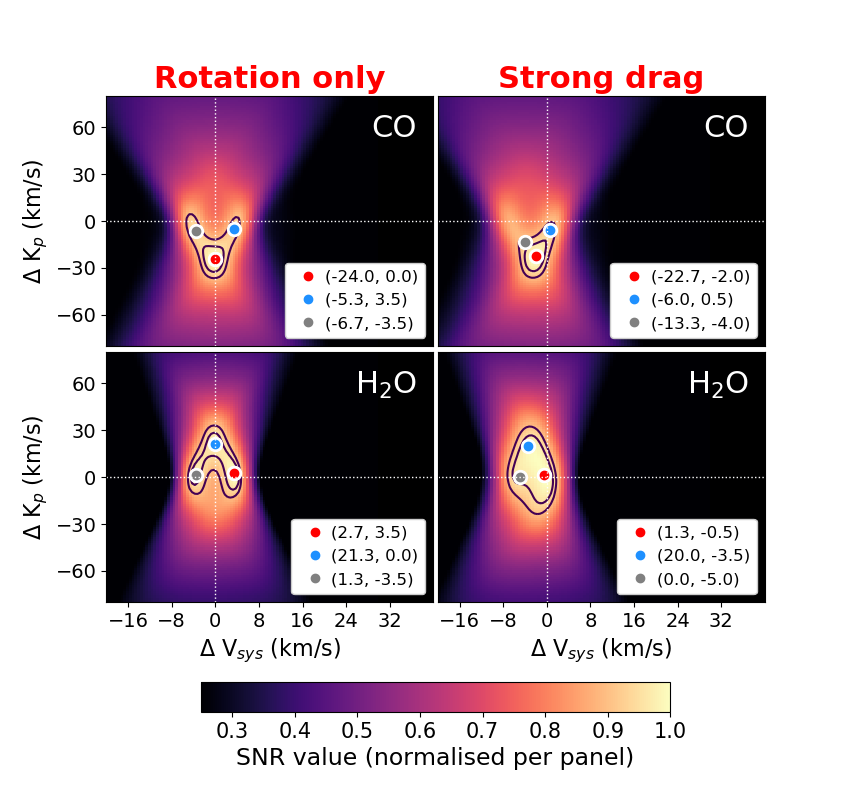}}
\vspace{-20pt}
\caption{{\color{black}{$K_\text{p}$--$V_\text{sys}$ maps of CO and H$_\text{2}$O for the nominal model with planet rotation only (i.e., winds switched off) and the nominal model with strong drag (i.e., $\tau_\text{drag} = 10^5$ s $\rightarrow$ $\tau_\text{drag} = 10^4$ s). The latter includes the effects of both winds and rotation. The red marker in each panel indicates the maximum SNR value \mbox{(SNR = 1)}, while the black contours correspond to SNR = 0.9 and SNR = 0.95, respectively. The red and grey markers show two other \mbox{($K_\text{p}$, $V_\text{sys}$)} combinations with high SNR values. The ($K_\text{p}$, $V_\text{sys}$) values associated with each of the markers are denoted in the legend (in units km/s).}}}
\label{fig:extra_kpvsys}
\end{figure}

\begin{figure}
\centering
\vspace{-10pt}
\makebox[\textwidth][c]{
\hspace{-250pt}
\includegraphics[width=0.54\textwidth]{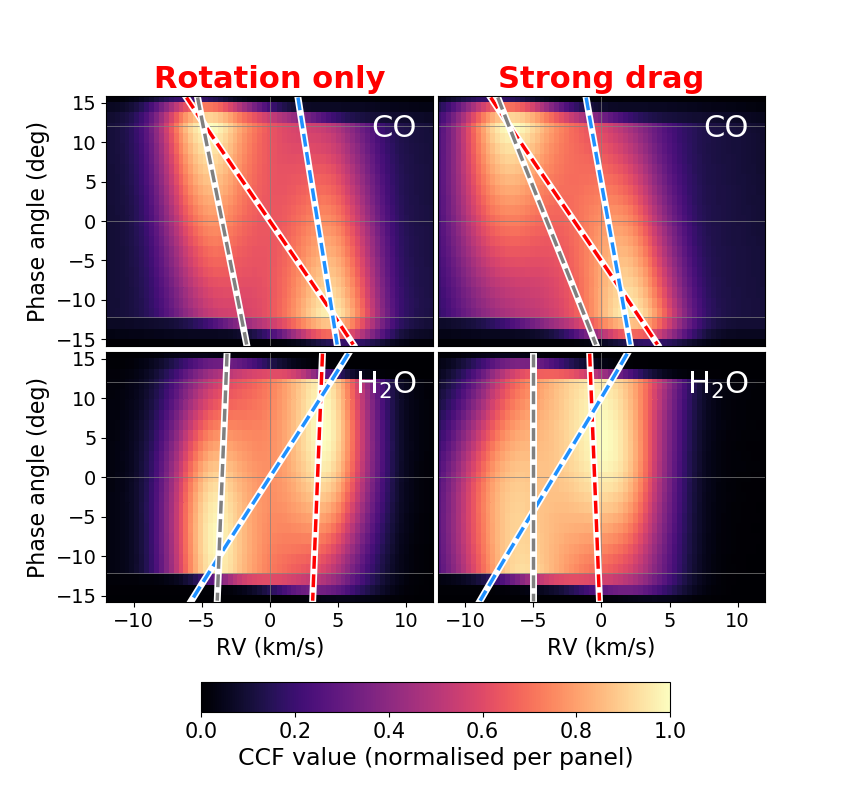}}
\vspace{-20pt}
\caption{{\color{black}{CCF maps of CO and H$_\text{2}$O for the ``rotation only'' and the ``strong drag'' versions of the nominal model (see also the right column of Fig. \ref{fig:extra_sector_plot}). In each panel, three planet trails are plotted. Each planet trail corresponds to the \mbox{($K_\text{p}$, $V_\text{sys}$)} values indicated by the marker with the same color in Fig. \ref{fig:extra_kpvsys}.}}}
\label{fig:extra_ccf}
\end{figure}

\subsubsection{The signature of planet rotation}

{\color{black}{The $K_\text{p}$--$V_\text{sys}$ maps in Fig. \ref{fig:all_kpvsys} (nearly) all show negative $K_\text{p}$ and $V_\text{sys}$ offsets. However, there are theoretical scenarios in which $\Delta K_\text{p}$ and/or $\Delta V_\text{sys}$ can be positive. To explore these scenarios, we revisit the two alternative versions of the nominal model presented in Fig. \ref{fig:extra_sector_plot}. Their  $K_\text{p}$--$V_\text{sys}$ maps are depicted in Fig. \ref{fig:extra_kpvsys}. The left panels show the SNR peaks of CO and water for the model with rotation only. In this scenario, the SNR peaks acquire a ``boomerang'' shape. That is, there is \emph{a family} of ($K_\text{p}$, $V_\text{sys}$) values that ``fit'' the absorption signal of the planet equally well. Fig. \ref{fig:extra_ccf} demonstrates why this is the case. For a model with rotation only, the absorption signal of the planet clearly features two components: a \emph{blueshifted} component associated with the trailing limb and a \emph{redshifted} component associated with the leading limb. Such a signal can be described by different trails that give rise to roughly the same integrated SNR: a trail fitting the trailing limb ($\Delta K_\text{p} \approx 0$; $\Delta V_\text{sys}<0$), a trail fitting the leading limb ($\Delta K_\text{p} \approx 0$; $\Delta V_\text{sys}>0$) and a trail that fits both components (large $K_\text{p}$ offset; $V_\text{sys}\approx0$). The latter has a negative slope for CO ($\Delta K_\text{p} < 0$), but a positive slope for water ($\Delta K_\text{p} > 0$), owing to the 3D distribution of these species across the atmosphere.

The right panels in Fig. \ref{fig:extra_kpvsys} show how the $K_\text{p}$--$V_\text{sys}$ maps of the planet change when weak winds are added to the model (the ``strong-drag'' scenario). Because of the presence of day-to-night winds, the planet signal becomes more blueshifted and the SNR peaks shift to negative $\Delta V_\text{sys}$. Also, the ``boomerang'' shape partly disappears, as winds tend to make the absorption trail of the planet smoother. However, especially for water, there are still a wide range of \mbox{($K_\text{p}$, $V_\text{sys}$)} combinations that fit the absorption signal of the planet well. }}

\subsubsection{Estimating the $K_\text{p}$ shift of a planet due to rotation}

In the scenario where planet rotation is dominating the Doppler shift of the planet, we can estimate the $K_\text{p}$ shift imposed on the planet signal. To this end, we assume that the absorption signal is dominated by the leading limb at the start of the transit and by the trailing limb at the end of the transit.

During transit, it holds that $\cos(\phi) \approx 1$, such that the change in radial velocity $\Delta$RV between two phases \emph{due to orbital motion} is 

\begin{equation}
    \Delta\text{RV} = 2 \pi K_\text{p} \bigg( \frac{\Delta\phi}{360^\circ}\bigg), 
\end{equation}

\noindent with $\Delta \phi$ the phase difference (in degrees). Therefore, \emph{in the planetary rest frame}, the $K_\text{p}$ shift resulting from planet rotation can be computed from 

\begin{equation}
    \Delta K_\text{p} = \frac{\Delta \text{RV}}{2\pi} \bigg( \frac{360^\circ}{\Delta\phi}\bigg), 
\end{equation}

\noindent with $\Delta \text{RV}$ the radial-velocity difference between the trailing and the leading limb, and $\Delta \phi$ the phase difference between ingress and egress. {\color{black}{The most extreme RV value that can be acquired by both limbs is $\pm v_\text{eq}$, the rotational velocity of the planet at the equator. Hence, a rough approximation\footnote{In reality, the average Doppler shift across the limb will be smaller than $v_\text{eq}$, as regions away from the equator lie closer to the rotation axis. Nonetheless, Fig. \ref{fig:extra_sector_plot} demonstrates that the assumption $\Delta \text{RV} = $ 2$v_\text{eq}$ is not too unrealistic, as the peaks of the CCFs of the full limb lie relatively close to $v_\text{eq}$ ($\pm$ 5.3 km/s). This is because the signal from the equatorial sectors is stronger than that of the polar sectors.} is $\Delta \text{RV} \approx \pm$2$v_\text{eq}$, resulting in

\begin{equation} \label{eq:max_kp_shift_1}
    \Delta K_\text{p} \approx \pm \frac{v_\text{eq}}{\pi} \bigg( \frac{360^\circ}{\Delta\phi}\bigg).
\end{equation}

\noindent Invoking $v_\text{eq} = 2\pi R_\text{p} / P$, $\Delta\phi = 2 \arcsin(R_*/a) \approx 2 R_*/a$, $360^\circ = 2\pi$ rad, and Kepler's third law, we can also write equation \ref{eq:max_kp_shift_1} as

\begin{equation} \label{eq:max_kp_shift_2}
    \Delta K_\text{p} \approx \pm \frac{R_\text{p}}{R_*} \bigg(\frac{ 2\pi GM_* }{P}  \bigg)^{1/3},
\end{equation}

\noindent with $R_\text{p}$ the planet radius, $P$ the orbital period, $a$ the semi-major axis of the orbit, $R_*$ the stellar radius, $M_*$ the stellar mass, and $G$ the gravitational constant, respectively. For signals dominated by the leading limb in the first half of the transit and by the trailing limb in the second half (e.g., Fe and CO), $\Delta K_\text{p}$ will be \emph{negative}. For signals dominated by the trailing limb in the first half of the transit and by the leading limb in the second half (e.g., H$_\text{2}$O), $\Delta K_\text{p}$ will be \emph{positive}. Hence, the sign of $\Delta K_\text{p}$ depends on the 3D distribution of a species across the atmosphere. 

Evaluating equation \ref{eq:max_kp_shift_2} for the parameters of WASP-76b\footnote{See e.g., \texttt{\href{http://exoplanet.eu/catalog/wasp-76_b/}{exoplanet.eu/catalog/wasp-76\_b/}}}, we find $\Delta K_\text{p} \approx$ $\pm$21 km/s, which is in rough agreement with the $K_\text{p}$ shifts reported for CO and water in Fig. \ref{fig:extra_kpvsys}. For WASP-121b, another well-studied ultra-hot Jupiter, we find $\Delta K_\text{p} \approx$ $\pm$28 km/s. This demonstrates that the $K_\text{p}$ offsets observed for a planet can be much larger than the actual line-of-sight velocities in its atmosphere.}}

\subsubsection{Comparison to transit observations}

{\color{black}{A considerable number of HRS observations of ultra-hot Jupiters have revealed peak offsets in $K_\text{p}$--$V_\text{sys}$ maps that hint at atmospheric dynamics and/or 3D spatial variations in temperature and chemistry.

\citet{Kesseli2022} and \citet{Pelletier2023} presented $K_\text{p}$--$V_\text{sys}$ maps for a plethora of species in the atmosphere of WASP-76b (H, Li, Na, Mg, K, Ca \textsc{ii}, V, Cr, Mn, Co, Ni, Sr \textsc{ii}, VO, Ca, Ba \textsc{ii}, O, Fe, and Fe \textsc{ii}). For the vast majority of these species, they reported \emph{negative} $K_\text{p}$ and $V_\text{sys}$ offsets, which is in good agreement with this work (see Fig. \ref{fig:all_kpvsys}). Note that many of the species observed in the optical are refractories and alkalis, which are abundant on the dayside of the planet. Therefore, their absorption signals should behave in the same way as those of iron, CO, and OH modelled in this work. For species such as H, O, and Ca \textsc{ii}, \citet{Kesseli2022} and/or \citet{Pelletier2023} found positive $K_\text{p}$ offsets. This is because the absorption lines of these species probe higher regions of the atmosphere that are likely prone to atmospheric escape (e.g., \citealt{Yan2021}). Such physics is not included in our model. 

In the infrared, CARMENES observations of WASP-76b revealed \emph{positive} $K_\text{p}$ offsets for H$_\text{2}$O, HCN ($+$50 km/s, \citealt{Sanchez-Lopez2022}), and OH ($+$35 km/s, \citealt{Landman2021}), suggesting a decreasing blueshift of the absorption lines over the course of the transit. For water, our models are able to produce a positive $\Delta K_\text{p}$ when the line-of-sight velocities are dominated by planet rotation (see Fig. \ref{fig:extra_kpvsys}). However, the expected offset would be of the order \mbox{$+20$ km/s} in this scenario. A +50 km/s shift in $K_\text{p}$ is hard to explain with our current framework. As for OH, our models predict $\Delta K_\text{p}$ to be negative, rather than positive. Further observations\footnote{The studies by \citet{Landman2021} and \citet{Sanchez-Lopez2022} were based on the same archival CARMENES data, so further observations would be needed to rule out the presence of any systematics in the dataset.} and/or modelling studies will be required to elucidate the differences between our models and the findings by \citet{Landman2021} and \citet{Sanchez-Lopez2022}.

Optical transmission observations of WASP-121b have shown \emph{negative} $K_\text{p}$ and $V_\text{sys}$ offsets for Fe (\citealt{Gibson2020, Borsa2021, Merritt2021}), Cr, V, Fe \textsc{ii} (\citealt{Borsa2021,Merritt2021}), Ca, K, Co, Cu, V \textsc{ii}, Ti \textsc{ii}, Mg, and Sc \textsc{ii} (\citealt{Merritt2021}). Many of the more ``exotic'' species reported by \citet{Merritt2021} only showed weak or tentative detections, so their ($K_\text{p}$, $V_\text{sys}$) values should be treated with caution. However, the observations demonstrate that the majority of refractories and alkalis undergo increasing blueshifts during the transit, just like on WASP-76b. \citet{Merritt2021} also reported a few species with $\Delta K_\text{p} \approx 0$ km/s and $\Delta V_\text{sys}<0$ km/s (Mn, Co \textsc{ii}, Ni), which could imply that these species are only observable on the trailing limb of the planet. More recently, \citet{Maguire2022} also recovered negative $K_\text{p}$ and $V_\text{sys}$ offsets when cross-correlating ESPRESSO data of WASP-121b with a template containing Fe, Mg, Cr, Ti, V, Na, and Ca lines. For Ca \textsc{ii}, \citet{Maguire2022} found a positive $\Delta K_\text{p}$, again indicating that its absorption lines probe higher regions of the atmosphere with different dynamics.

For KELT-20b/MASCARA-2b, \citet{Nugroho2020} and \citet{Rainer2021} reported ``double-peak'' features in the $K_\text{p}$--$V_\text{sys}$ maps of neutral iron. These could hint at the fact that planet rotation is the dominant contributor to the line-of-sight velocities, such that the absorption signal is made up of separate components associated with the trailing and leading limb, respectively (as in Figs. \ref{fig:extra_kpvsys} and \ref{fig:extra_ccf}). What is puzzling however, is that the SNR peaks lie 70--80 km/s apart along the $K_\text{p}$ axis, while equation \ref{eq:max_kp_shift_2} only predicts $\Delta K_\text{p} \approx 20$ km/s for KELT-20b. Furthermore, \citet{Rainer2021} observed five transits of KELT-20b, and only in two transits was the double-peak feature recovered.     

Other transiting ultra-hot Jupiters for which peak offsets in $K_\text{p}$--$V_\text{sys}$ maps were found are WASP-189b (\citealt{Prinoth2022}), HAT-P-70b (\citealt{Bello-Arufe2022}), and KELT-9b (\citealt{Borsato2023}). For 18 species present in the atmosphere of KELT-9b, \citet{Borsato2023} extracted $K_\text{p}$ values spanning a range of 60 km/s (see their Fig. 6). Interpreting their observations with our current set of models is hard, as the equilibrium temperature of KELT-9b is roughly two times that of WASP-76.}}

\subsubsection{A note on high-resolution retrievals of ultra-hot Jupiters}

{\color{black}{In this work, we showed that different species are subject to different Doppler shifts and $K_\text{p}$--$V_\text{sys}$ offsets in transmission. At the moment, retrieval frameworks typically include one $\Delta K_\text{p}$ and one $\Delta V_\text{sys}$ parameter to describe the ``bulk'' Doppler shift of the entire spectrum as a function of phase (\citealt{Gandhi2022,Gandhi2023,Maguire2022,Pelletier2023}). In the optical, such an approach is justified for the vast majority species (i.e., most alkalis and refractories) as these are expected to have similar distributions across the atmosphere, resulting in similar $K_\text{p}$--$V_\text{sys}$ offsets (e.g., Figs. 1 and 9 in \citealt{Pelletier2023}). However, as noted by \citet{Maguire2022} and \citet{Pelletier2023}, care should be taken with species that probe the planet's exosphere (e.g., H, O, Fe \textsc{ii}, Mg \textsc{ii}, and Ca \textsc{ii}). Both studies excluded these species from their retrievals, as the exosphere is non-hydrostatic (impacting line strengths) and features strong outflows (impacting line shapes and positions). A 1D retrieval model with \emph{one} set of \mbox{($\Delta K_\text{p}$, $\Delta V_\text{sys}$)} parameters and a single scale factor (a parameter controlling the line strengths of the model) cannot account for the behaviour of all species at the same time. 

Following \citet{Pelletier2023}, a good practice for high-resolution retrievals would be to plot the $K_\text{p}$--$V_\text{sys}$ maps of all species to be included in the forward model, and examine their peak offsets. If the peak offsets of two (groups of) species are substantially different, they may require their own set of ($\Delta K_\text{p}$, $\Delta V_\text{sys}$) parameters. Another, option is to run a separate retrieval for each (group of) species. The latter does not increase the complexity of the forward model, but doubles the computing time. 

In the infrared, things are more intricate than in the optical as water and CO -- the two most prominent species -- probe completely different parts of the atmosphere (see Fig. \ref{fig:absorption_regions_all_1}), each with their own temperature, dynamics, and scale height. Therefore, fitting the same $\Delta K_\text{p}$, $\Delta V_\text{sys}$, and temperature profile to the absorption lines of both species may be problematic. The most straightforward solution would be to run two separate retrievals for water and CO.

Contrary to CO, which probes the dayside during the entire transit, the absorption regions of water shift across the terminator as a function of orbital phase. On the trailing limb, the absorption regions shift from the nightside to the dayside, while they shift from the dayside to the nightside on the leading limb. Therefore, water would be the ideal molecule to study with a 2D retrieval model (\citealt{Gandhi2022,Gandhi2023}), which is able to assign separate temperatures and abundances to the trailing and leading limb of the planet, respectively. }}

\section{Conclusion}
\label{sec:conclusion}

Developing a deeper understanding of the ``3D-ness'' of exoplanet atmospheres is crucial to fully leverage the information content of both their high-resolution and low-resolution spectra. With JWST delivering its first data (e.g., \citealt{JWST2022}) and a new generation of ground-based telescopes (E-ELT, GMT, TMT) on the horizon, modelling studies that bridge the gap between theory and observation play an essential role in the interpretation of current and future observations. In this work, we simulated the cross-correlation signals of Fe, CO, H$_2$O, OH, and TiO for four different 3D models of a benchmark ultra-hot Jupiter (WASP-76b) in transmission. Because ultra-hot Jupiters show extreme spatial variations in temperature and chemistry across their terminators, their transmission spectra contain a wealth of information about the 3D structure of the atmosphere. 

VLT/ESPRESSO and GEMINI-N/MAROON-X are able to phase-resolve the absorption signals of ultra-hot Jupiters in the optical (\citealt{Ehrenreich2020,Borsa2021,Pelletier2023}). With novel spectrographs such as GEMINI-S/IGRINS (\citealt{Mace2018,Line2021}) and VLT/CRIRES$+$ (\citealt{Holmberg2022,Dorn2023}), this will now also possible in the infrared. Moreover, once the E-ELT is on sky, phase-resolving the CCF will become standard practice for any high-resolution observation of a hot gas giant, as the signal-to-noise will be high enough to detect the planet in only a fraction of a transit. Also, the E-ELT will offer the opportunity to take ingress and egress spectra, whereby only a part of the planet disk is blocking the star. 

We summarise our most important findings below:

\begin{enumerate}
   \setlength\itemsep{1em}


   \item[$\bullet$] For species that probe the \emph{dayside} of an ultra-hot Jupiter (refractories like Fe, or stable molecules like CO and OH), the net blueshift should increase during the transit, resulting in a \emph{negative} $K_\text{p}$ offset. This holds even in the absence of an east-west asymmetry (e.g., due to a hotspot offset). The increasing blueshift is due to the combined effect of the 3D spatial distribution of the species and planet rotation. {\color{black}{Our findings are in good agreement with optical high-resolution observations of WASP-76b and WASP-121b (e.g., \citealt{Merritt2021,Borsa2021,Kesseli2022,Maguire2022,Pelletier2023}). Conversely, for species that probe the \emph{nightside} (such as H$_\text{2}$O and TiO), their 3D spatial distribution and planet rotation act in an opposite manner. Depending on the 3D wind profile of the planet, this can lead to weaker blueshifts with orbital phase, or even increasing redshifts. Such behaviour results in a \emph{positive} $K_\text{p}$ offset.}} 

    \item[$\bullet$] {\color{black}{The $K_\text{p}$ offset of a species reflects the \emph{rate of change} of its Doppler shift in the planetary rest frame. Therefore, as opposed to $\Delta V_\text{sys}$ (which is of the same order as the wind speeds), $\Delta K_\text{p}$ can be much larger than the line-of-sight velocities in the planet's atmosphere at any time. $\Delta K_\text{p} < 0$ when the Doppler shift becomes more negative during the transit, while \mbox{$\Delta K_\text{p} > 0$} when the Doppler shift becomes more positive. In this work, we derived a formula to estimate the typical $K_\text{p}$ offset of a planet. For \mbox{WASP-76b} and \mbox{WASP-121b}, $\Delta K_\text{p}$ can be as large as $\pm 21$ km/s and \mbox{$\pm 28$ km/s}, respectively.}} 

    \item[$\bullet$] {\color{black}{When performing atmospheric retrievals on transmission spectra of ultra-hot Jupiters, \emph{separate} temperature profiles and \mbox{($\Delta K_\text{p}$, $\Delta V_\text{sys}$)} values should be retrieved for species that probe the dayside and the nightside of the atmosphere, respectively (e.g., CO and H$_\text{2}$O in the infrared). Our analytical formula can provide a reasonable prior for the range of possible departures from a planet's orbital $K_\text{p}$ value.}}

    \item[$\bullet$] {\color{black}{For WASP-76b, our nominal GCM model does not predict strong differences between the cross-correlation signals of Fe, CO, H$_\text{2}$O and OH in transmission. However, our model with a colder morning limb, which produces the same ``kink'' feature as seen in the data ( \citealt{Ehrenreich2020,Kesseli2021,Pelletier2023}), predicts a more diverse set of absorption signals for the chemical species studied. We conclude that observing the phase-dependent absorption signal of \emph{multiple} species that probe distinct parts of the atmosphere allows to differentiate between two models that fit the signal of a single species equally well.}} 

    \item[$\bullet$] Even though CO is uniformly distributed across the atmosphere of an ultra-hot Jupiter, it predominantly probes the dayside. This is because of a ``shielding effect''. Since the dayside is more extended than the nightside, CO absorption happens at high altitudes on the dayside where the nightside contribution to the optical depth is zero.  

    \item[$\bullet$] H$_2$O absorption lines can be strongly muted by optically-thick clouds on the nightside of ultra-hot Jupiters. On the other hand, nighside clouds will not have a big impact on the absorption signals of Fe and CO, as these species probe higher altitudes on the dayside.

\end{enumerate}

\section*{Acknowledgements}

We are grateful to Ray Pierrehumbert for sharing computing resources. We also thank David Ehrenreich and Ray Pierrehumbert for insightful discussions. JPW sincerely acknowledges support from the Wolfson Harrison UK Research Council Physics Scholarship and the Science and Technology Facilities Council (STFC). This work benefited from the 2022 Exoplanet Summer Program in the Other
Worlds Laboratory (OWL) at the University of California, Santa Cruz, a program funded by the Heising-Simons Foundation. Finally, we thank the anonymous referee for thoughtful comments that helped improve the quality of the manuscript.

\section*{Data Availability}

The data and models underlying this article will be shared on reasonable request to the corresponding author.



\bibliographystyle{mnras}
\bibliography{citations} 



\appendix

\section{Impact of new modelling approaches on the CCF map}
\label{ap:A}

\begin{figure*}
\centering
\vspace{-20pt}
\makebox[\textwidth][c]{\hspace{30pt}\includegraphics[width=1.1\textwidth]{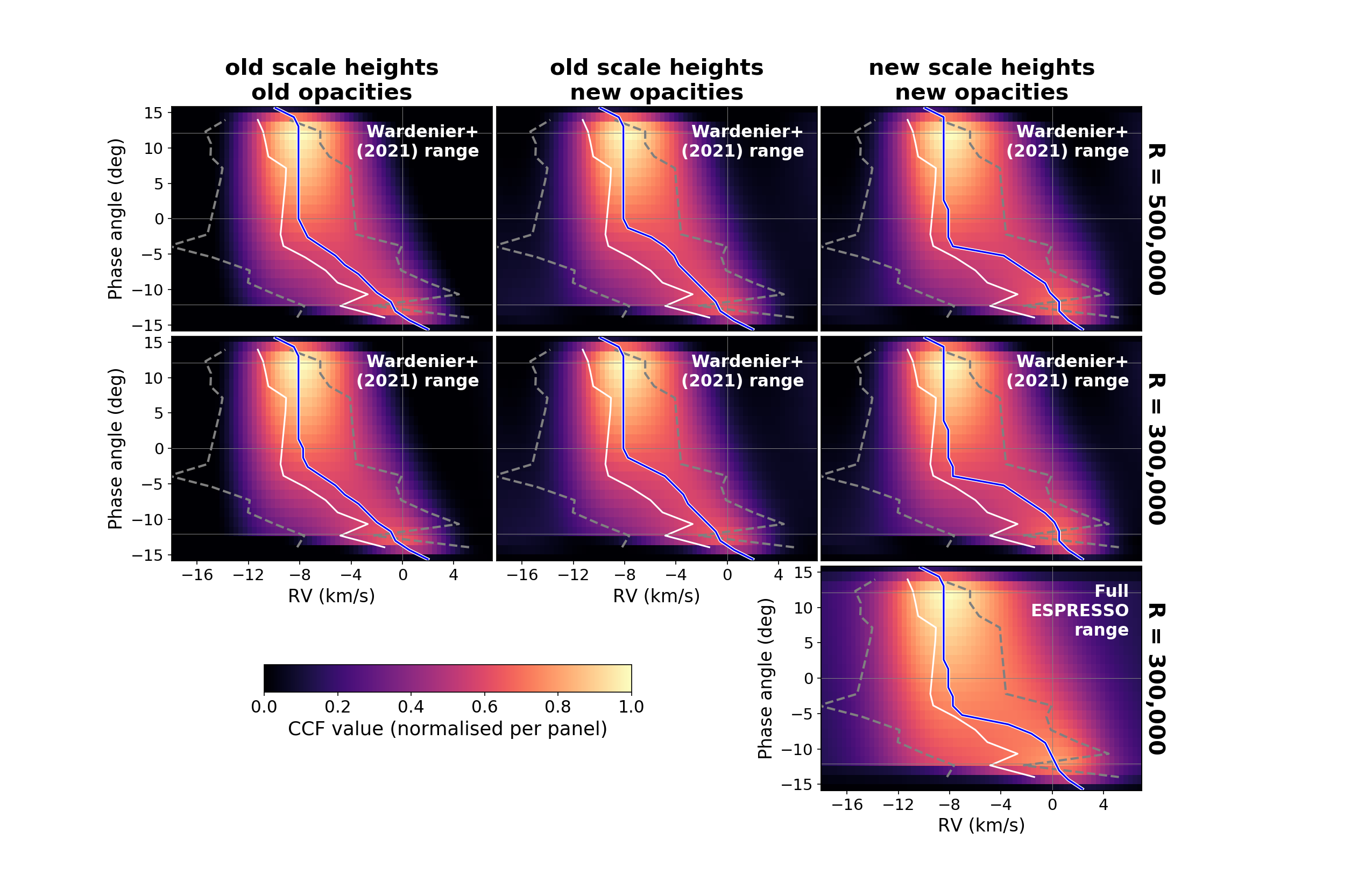}}
\vspace{-30pt}
\caption{CCF maps for iron (Fe \textsc{i}) in the cold-morning-limb model. The panel in the top left shows the CCF map as presented in \citet{Wardenier2021}, while the panel in the bottom right shows the CCF map computed for \emph{the same atmosphere} in this work. The other panels illustrate how the CCF map is impacted by each of the four ``modelling changes'' discussed in Section \ref{subsec:optical_spectra}. In each panel, the blue curve indicates the maximum value of the CCF at each orbital phase. The white curve is the iron signal of WASP-76b observed by \citet{Ehrenreich2020}. Grey dashes indicate the FWHM of this signal.}
\label{fig:appendix_plot}
\end{figure*}

Fig. \ref{fig:appendix_plot} shows a comparison between the CCF map of iron obtained for the cold-morning-limb model in \citet{Wardenier2021} (top left) and the cold-morning-limb model from this work (bottom right). The underlying atmosphere is the same, but a few changes were made to the radiative transfer: (i) accounting for scale-height differences due to hydrogen dissociation, (ii) including opacities for more species, and using iron line lists with pressure broadening and no line-wing cut-off, (iii) increasing the wavelength range, and (iv) decreasing the resolution (see Section \ref{subsec:optical_spectra}).

In summary, we find that changes (iii) and (iv) have the biggest impact on the CCF map. However, the overall behaviour of the iron signal proves robust -- an increasing blueshift and signal strength over the course of the transit, with the blueshift remaining constant at about $-8$ km/s after mid-transit. 

\bsp	
\label{lastpage}
\end{document}